\documentclass[superscriptaddress,prb,twocolumn,notitlepage,showkeys,showpacs,longbibliography]{revtex4-1}
\usepackage[utf8]{inputenc}
\usepackage{stmaryrd}
\usepackage{amsmath}
\usepackage{amssymb}
\usepackage{graphicx}
\usepackage{dcolumn}
\usepackage{bm}
\usepackage{multirow}
\usepackage[colorlinks,linkcolor=blue,citecolor=blue,urlcolor=blue]{hyperref}
\usepackage{color}
\usepackage{ulem}
\usepackage{comment}
\usepackage{booktabs}
\usepackage{verbatim}
\newcommand{\etal}{\textit{et al.}}

\begin{document}
\title{
Emergent topological quantum orbits in the charge density wave phase of kagome metal CsV$_3$Sb$_5$
}
\author{Hengxin Tan}
\author{Yongkang Li}
\author{Yizhou Liu}
\author{Daniel Kaplan}
\affiliation{Department of Condensed Matter Physics, Weizmann Institute of Science, Rehovot 7610001, Israel}
\author{Ziqiang Wang}
\affiliation{Department of Physics, Boston College, Chestnut Hill, Massachusetts 02467, USA}
\author{Binghai Yan}
\email{binghai.yan@weizmann.ac.il}
\affiliation{Department of Condensed Matter Physics, Weizmann Institute of Science, Rehovot 7610001, Israel}

\begin{abstract}
The recently discovered kagome materials $A$V$_3$Sb$_5$ ($A$ = K, Rb, Cs) attract intense research interest in intertwined topology, superconductivity, and charge density waves (CDW). Although the in-plane $2\times2$ CDW is well studied, its out-of-plane structural correlation with the Fermi surface properties is less understood. In this work, we advance the theoretical description of quantum oscillations and investigate the Fermi surface properties in the three-dimensional CDW phase of CsV$_3$Sb$_5$. We derived Fermi-energy-resolved and layer-resolved quantum orbits that agree quantitatively with recent experiments in the fundamental frequency, cyclotron mass, and topology.
We reveal a complex Dirac nodal network that would lead to a $\pi$ Berry phase of a quantum orbit in the spinless case. However, the phase shift  of topological quantum orbits is contributed by the orbital moment and Zeeman effect besides the Berry phase in the presence of spin-orbital coupling (SOC). Therefore, we can observe topological quantum orbits with a $\pi$ phase shift in otherwise trivial orbits without SOC, contrary to common perception.
Our work reveals the rich topological nature of kagome materials and paves a path to resolve different topological origins of quantum orbits.
\end{abstract}

\maketitle

\section{Introduction}

The recently discovered kagome superconductors $A$V$_3$Sb$_5$ ($A$ = K, Rb, Cs) \cite{Ortiz2019new} stimulate extensive studies for their intriguing charge density waves (CDW) \cite{jiang2021unconventional,Ortiz2020Z2,tan2021charge,chen2021roton,Liang2021three,Park2021,zhao2021cascade,Christensen2021theory,Denner2021analysis}, possible symmetry-breaking without magnetism \cite{yang2020giant,jiang2021unconventional,Kenney2021absence,mielke2022time,yu2021evidence,Li2022no,Feng2021chiral,Feng2021low,Khasanov2022time,xu2022universal,saykin2022high},
topological band structure \cite{Ortiz2020Z2,tan2021charge,hu2022topological,Hu2022rich,kang2022twofold}, and exotic superconductivity\cite{Ortiz2021superc,Yin2021superc,chen2021roton,Chen2021double}.
Aiming at deriving the Fermi surface (FS) property of CsV$_3$Sb$_5$, a dozen quantum oscillation experiments reveal complicated but similar oscillation frequencies \cite{yu2021concurrence,Ortiz2021fermi,Fu2021quantum,Gan2021magneto,Shrestha2022nontrivial,Chen2022anomalous,zhang2022emergence,chapai2022magnetic,Broyles2022the},
bearing the Fermi energy variation in different samples.
However, the non-trivial Fermi pockets (or quantum orbits), characterized by a $\pi$ phase shift in the fundamental quantum oscillation, are dissimilar in different reports.
For example, Fu $\etal$ \cite{Fu2021quantum} claimed nontrivial pockets are 73 T and 727 T in the fundamental oscillation frequency.
Chapai $\etal$ \cite{chapai2022magnetic} found Fermi pockets of 79, 736, and 804 T are non-trivial.
Broyles $\etal$ \cite{Broyles2022the} reported 28, 74, and 85 T to have a $\pi$ phase shift.
Shrestha $\etal$ \cite{Shrestha2022nontrivial} stated that all seven quantum orbits (from 18 to 2135 T) observed are topological.

It is commonly believed that the measured $\pi$ phase shift in a quantum oscillation originates from the $\pi$ Berry phase ($\phi_B$) of Dirac fermion in the system \cite{Mikitik,Lukyanchuk2004phase}.
For example, graphene exhibits the $\pi$ Berry phase in quantum oscillation \cite{zhang2005experimental}.
However, as pointed out by Alexandradinata $\etal$ \cite{Aris2018PRX,Aris2018PRB}, the $\pi$ phase shift in the quantum oscillation
has broader origins than the $\pi$ Berry phase of a Dirac fermion because of the orbital magnetic moment and Zeeman effect.
The success of manifesting the Berry phase of graphene lies in the negligible spin-orbital coupling (SOC) \cite{Min2006intrinsic,Yao2007spin}, resulting in a zero orbital magnetic moment-induced phase $\phi_R$ and a small Zeeman effect-induced phase $\phi_Z$ as a spin reduction factor in the quantum oscillation intensity.
In spinful systems (the SOC of $A$V$_3$Sb$_5$ is about dozens of meV), the non-zero $\phi_R$ and $\phi_Z$ due to spin-momentum locking sensitive to the Fermi energy cannot be separated from $\phi_B$ in the total phase shift.
As a result, the SOC-driven band anti-crossing can generate quantum orbits with a $\pi$ phase shift in a proper energy window.
Thus, the origin of observed non-trivial Fermi pockets of CsV$_3$Sb$_5$ may deviate from the vaguely argued $\pi$-Berry phase mechanism, which remains an open question. Furthermore, the dissimilar frequencies of non-trivial Fermi pockets of CsV$_3$Sb$_5$ remain to be reconciled.

Besides the discrepancy in the topological nature of Fermi pockets, another intriguing puzzle is their structural origin.
Previous calculations based on the 2$\times$2$\times$1 CDW model of CsV$_3$Sb$_5$ cannot rationalize their respective quantum oscillation experiments\cite{Ortiz2021fermi,Fu2021quantum}.
The 2$\times$2$\times$1 CDW, i.e., the star of David (SD) and inverse star of David (ISD)\cite{tan2021charge} structures, have been commonly adopted to understand various experiments and conceive theoretical models.
Recent experiments suggest two-fold or four-fold interlayer modulation with alternative SD and ISD layers \cite{Ortiz2021fermi,Liang2021three,Li2021observation,song2022orbital,Stahl2022temperature,Hu2022Coexistence,kang2022charge}.
The interlayer structural modulation-induced FS reconstruction, which may help to understand the emergence of novel density wave orders and superconductivity at lower temperatures \cite{zhou2021chern}, remains elusive.

In this work, we resolved the origin and topology of the observed quantum orbits in CsV$_3$Sb$_5$ from theory for the first time.
We derived Fermi-energy-resolved and layer-resolved quantum orbits that agree quantitatively with experiments in the cyclotron frequency and non-trivial phase shift.
According to the layer contribution, we classified all quantum orbits into three groups, i.e., SD, ISD, and mixed (SD+ISD) group, where most small quantum orbits show a clear 3D nature.
According to the topological origin, the non-trivial quantum orbits are classified into four types, with the topology of one type originating from the $\pi$ Berry phase and the other three originating from the SOC jointly with the Zeeman effect.
Most importantly, the SOC and/or Zeeman effect can lead to a trivial quantum orbit (zero phase shift in the quantum oscillation), even though the orbit has a $\pi$ Berry phase in the spinless case.
A hidden Dirac nodal network protected by mirror symmetries in the 3D CDW phase is revealed, which is weakly gapped in the presence of SOC \cite{Burkov2011,Fang2015}.
We not only resolve the topology of quantum orbits but also reveal the appreciable 3D nature of Fermi surfaces in CsV$_3$Sb$_5$.


\section{Results and discussions}

\begin{figure*}[tbp]
\includegraphics[width=0.7\linewidth]{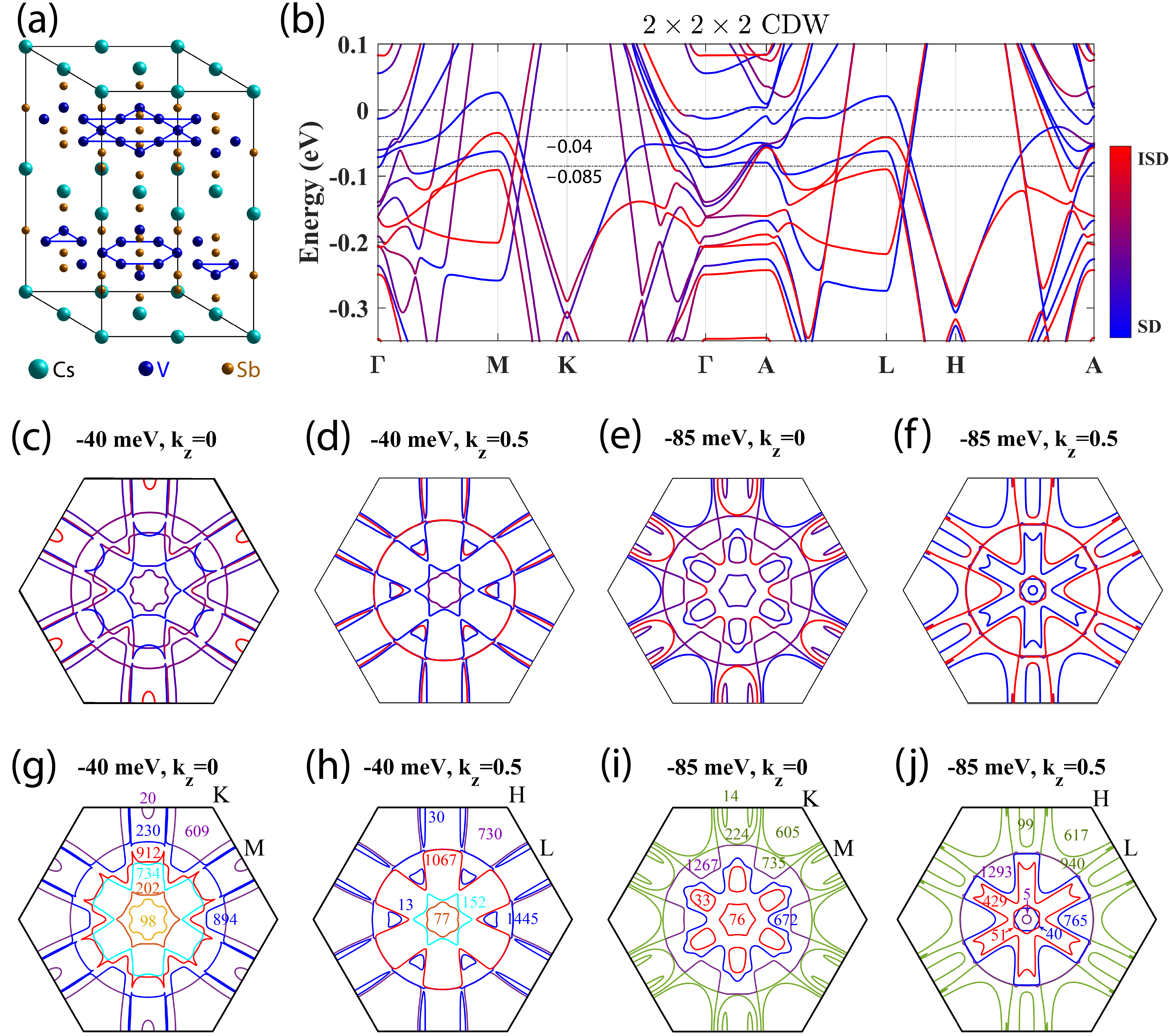}
\caption{\label{Fig-band-FS} \textbf{Crystal structure and Layer-resolved band structure and Fermi surfaces of the 2$\times$2$\times$2 CDW.}
(a) The crystal structure with alternating SD and ISD layers.
(b) Band structure. The color bar indicates the ISD (red) and SD (blue) contributions. The energy of the charge neutral point is set to zero. Energies of $-$40 and $-$85 meV are indicated by dash-dotted lines.
(c)-(f) Layer-resolved Fermi surfaces in the first Brillouin zone (BZ) of the CDW state at $-$40 meV and $-$85 meV on $k_z=0$ and 0.5 ($2\pi/c$) planes. The color bar is the same as (b).
(g)-(j) show cyclotron frequencies (in unit of T) of the corresponding Fermi surfaces where the line color differentiates different Fermi surfaces.
These seemingly open Fermi surfaces at the BZ boundary extend to neighboring BZs and form closed Fermi pockets. SOC is included in calculations.
}
\end{figure*}

We consider three hexagonal CDW structures with the $P6/mmm$ space group symmetry, i.e., the 2$\times$2$\times$1 SD and ISD and the 2$\times$2$\times$2 CDW. The energetic stability of interlayer stacking has been theoretically studied\cite{tan2021charge,Park2021,Christensen2021theory,Ortiz2021fermi,Subedi2022hexagonal}. We adopt the 2$\times$2$\times$2 CDW alternating SD and ISD layers without interlayer lateral shift as suggested by recent APRES measurements \cite{kang2022charge,Hu2022Coexistence}, shown in Fig. \ref{Fig-band-FS}(a). As we will see, the 2$\times$2$\times$2 model captures the essential interlayer interaction which also occurs in more complicated structures like the 2$\times$2$\times$4 CDW \cite{Ortiz2021fermi}.

\begin{figure*}[tbp]
\includegraphics[width=0.8\linewidth]{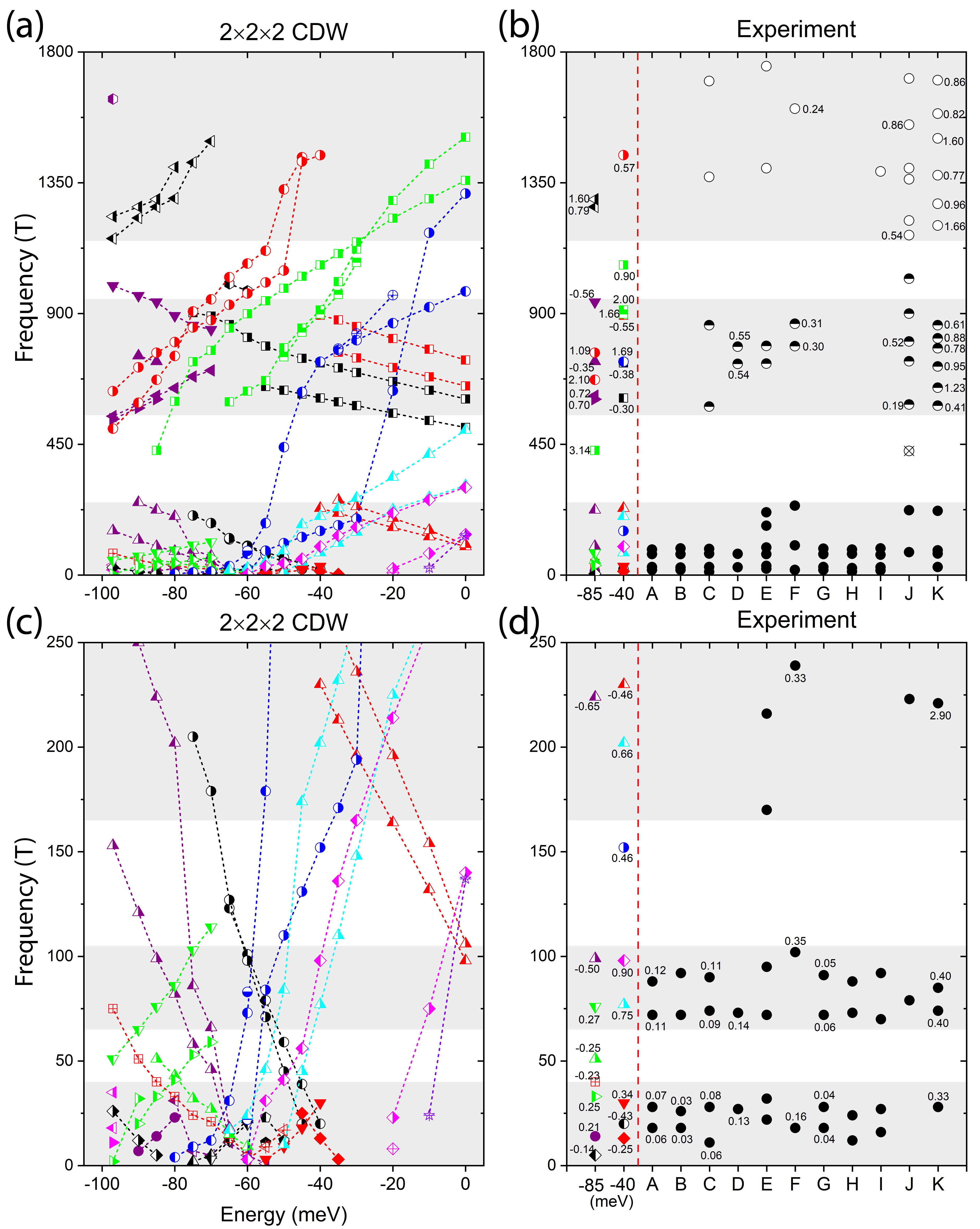}
\caption{\label{Fig-freq} \textbf{Calculated cyclotron frequencies in comparison with experiments.}
(a) Cyclotron frequencies of extremal quantum orbits as a function of Fermi energy for the 2$\times$2$\times$2 CDW. The energy of the charge neutral point is set to energy zero. Symbols with the same color and style connected by dashed lines represent the evolution of the same quantum orbit.
(b) Comparison of calculated cyclotron frequencies at $-$40 and $-$85 meV (on the left side of the red dashed line) with experimental values in literature.
Experimental set A is from Ref. \onlinecite{Gan2021magneto}, B from Ref. \onlinecite{yu2021concurrence}, C from Ref. \onlinecite{Ortiz2021fermi}, D from Ref. \onlinecite{Fu2021quantum}, E and F from Ref. \onlinecite{Shrestha2022nontrivial} for the angle between magnetic filed and $c$ axis being 0$^{\circ}$ and 20$^{\circ}$, respectively, G from Ref. \onlinecite{Chen2022anomalous}, H and I from Ref. \onlinecite{zhang2022emergence} for bulk and flake, respectively, J from Ref. \onlinecite{chapai2022magnetic}, and K from Ref. \onlinecite{Broyles2022the}.
Numbers marked near the symbols are corresponding cyclotron masses in unit of $m_{0}$.
The experimental values are divided into three regions, i.e., low-frequency region 0$\sim$250 T, medium-frequency region 550$\sim$950 T, and high-frequency region $\geq$1150 T.
(c) and (d) are a zoom-in of the low-frequency region in (a) and (b), respectively, which can be further divided into three sub-regions as indicated by the three grey backgrounds.
}
\end{figure*}

The band structure calculated by density-functional theory (DFT) is shown for the 2$\times$2$\times$2 CDW in Fig. \ref{Fig-band-FS}(b). By layer resolution, bands and FSs are well separated into SD, ISD, and mixed SD+ISD types. The band structure can be regarded as overlapping SD and ISD phases with considerable hybridization (see Supplementary Fig. S1). Because the chemical potential in experiments \cite{Ortiz2021fermi,yang2020giant,kang2022twofold,Hu2022rich,Luo2022electronic} is usually lower than the theoretical charge neutral point (energy zero), we will focus on the Fermi energy window of $-100$ to 0 meV and investigate corresponding FS properties. We show two typical Fermi energies at --40 and --85 meV in Fig. \ref{Fig-band-FS}(c)-(j) to demonstrate the layer- and energy-dependence of multiple quantum orbits.

We obtain extremal quantum orbits by extracting the maximal and minimal FS area along the $k_z$ axis for the 2$\times$2$\times$2 CDW and show their Fermi energy dependence in Fig. \ref{Fig-freq}(a)\&(c). Most extremal orbits distribute in high-symmetry planes of $k_z = 0 / 0.5$ (in unit of 2$\pi/c$, $c$ is the out-of-plane lattice parameter) while some special cases appear at a generic $k_z$ (see an example at --40 meV in Table \ref{Table_freq}). We will identify the proper Fermi energy by comparing calculated cyclotron frequencies with experimental values.

We summarized exhaustively recent quantum oscillation measurements \cite{yu2021concurrence,Ortiz2021fermi,Fu2021quantum,Gan2021magneto,Shrestha2022nontrivial,Chen2022anomalous,zhang2022emergence,chapai2022magnetic,Broyles2022the} in Fig. \ref{Fig-freq}(b)\&(d). Most experimental results can be divided into three groups, i.e., the low-frequency region 0$\sim$250 T, medium-frequency region 550$\sim$950 T, and high-frequency region $\geq$ 1150 T. Many high frequencies may be induced by the magnetic breakdown by merging neighboring FSs \cite{Aris2018PRB}. Thus, we will focus on cyclotron frequencies no larger than 1800 T. There are 4$\sim$6 cyclotron frequencies in the low-frequency region, 2$\sim$6 frequencies in the medium-frequency region, and 1$\sim$6 frequencies in the frequency region of 1150$\sim$1800 T.

Considering both the number and distribution of frequencies in three regions, we find theoretical results at energy ranges of [$-$50, $-$35] meV and [$-$90, $-$75] meV match well with most experiments. One may also find results in the range of [$-$20, 0] meV match experiments at medium- and high-frequency regions. However, the agreement in the low-frequency region in this energy range is poor [see Fig. \ref{Fig-freq}(c)-(d)]. In the following, we will analyze two representative Fermi energies, $-$40 and $-$85 meV, as shown in Fig.~\ref{Fig-band-FS} and Table \ref{Table_freq}. In detail, the 2$\times$2$\times$2 CDW shows 8 frequencies in the low-frequency region and 7 frequencies in the medium-frequency region at both $-$40 and $-$85 meV. In the high-frequency region, it shows 1 frequency at $-$40 meV and 2 frequencies at $-$85 meV. In general, theoretical results in low- and medium-frequency regions agree better with experiments than the high-frequency region, as shown in Fig. \ref{Fig-freq}. The highest frequencies at $-$40 or $-$85 meV correspond to nearly isotropic FSs centered at $\Gamma$ (see Fig. \ref{Fig-band-FS}), which originates in Sb-$p_z$ bands. Corresponding FSs are very close to several neighboring pockets. Thus, magnetic breakdown may occur to generate even larger frequencies in experiments, rationalizing the magnetic breakdown observed in Ref. \onlinecite{chapai2022magnetic}.

\begin{figure}[tbp]
\includegraphics[width=0.95\columnwidth]{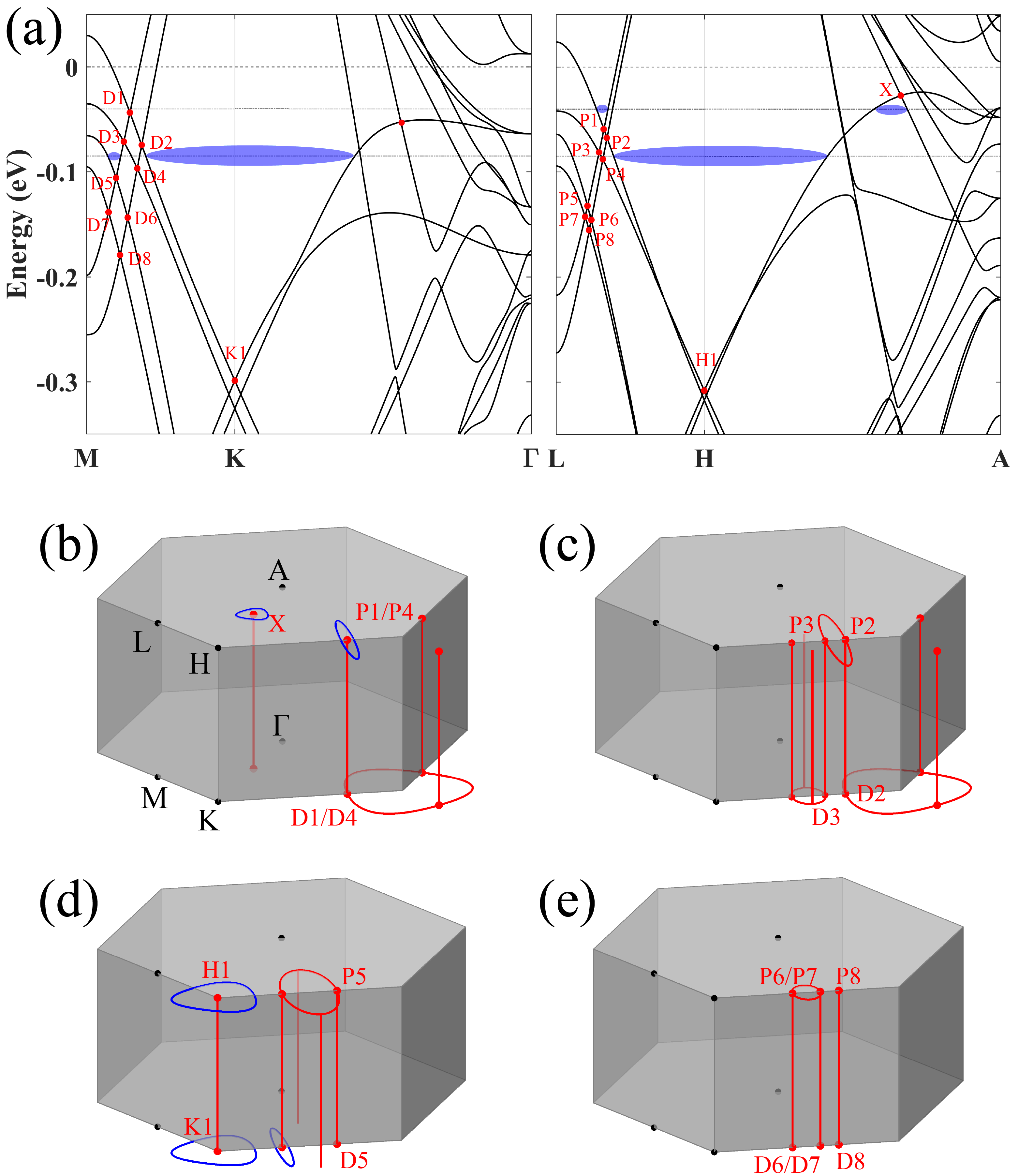}
\caption{\label{Fig4-Dirac} \textbf{Dirac nodal lines and nodal networks in the band structure. }
(a) Band structures on $k_z = 0$ (left part) and 0.5 (right part) planes without SOC. Eight Dirac points on the $MK$ ($LH$) line near the charge neutral point are labeled from D1 to D8 (P1--P8) respectively. Selected Dirac points at $K$ and $H$ are labeled as $K1$ and $H1$, respectively.
The two dash-dotted horizontal lines indicate the energy positions of $-$40 and $-$85 meV, respectively, where non-trivial Fermi surfaces ($\phi_B=\pi$) are indicated by blue ellipses.
Dirac nodal lines/rings connecting D1 to D8 (P1--P8) Dirac points are divided into four groups, as sketched in (b)-(e). Blue circles in (b) and (d) indicate Fermi surfaces surrounding a nodal line at $-$40 and $-$85 meV in (a)\&(b), respectively.
For simplicity, we show those Dirac nodal lines/rings and Fermi surfaces partly near one $M/L$ and one $K/H$ point. The full Dirac nodal lines/rings and non-trivial Fermi surfaces can be obtained by crystal symmetries ($P6/mmm$). We note that nodal lines except the $K1-H1$ line are not necessarily straight along $k_z$ but locate inside the mirror plane.
}
\end{figure}

In contrast, we cannot find a satisfactory agreement between experiments and the 2$\times$2$\times$1 CDW (SD or ISD) simultaneously in all three frequency regions. The quantitative comparison similar to Fig.~\ref{Fig-freq} is shown in Supplementary Fig. S2. Our results are consistent with Refs. \onlinecite{Ortiz2021fermi,Fu2021quantum} where the SD or ISD could not match all frequencies in their respective experiments.

Furthermore, we compare with experiments the cyclotron mass $m^*$ (in unit of bare electron mass $m_{0}$) of quantum orbits in Table \ref{Table_freq} and Fig. \ref{Fig-freq}.
In the low-frequency region in Fig. \ref{Fig-freq}(d), reported masses are rather diverse in literature.
Several experiments found $m^*=$ 0.03$\sim$0.14 while Shrestha $\etal$ \cite{Shrestha2022nontrivial} and Broyles $\etal$ \cite{Broyles2022the} found $m^* =$ 0.16$\sim$0.40, [sets F and K in Fig. \ref{Fig-freq}(d)].
Our calculations agree better with Shrestha $\etal$ \cite{Shrestha2022nontrivial} and Broyles $\etal$ \cite{Broyles2022the} (except the large value 2.90 for the observed frequency 221 T) in the low-frequency region.
In the medium- and high-frequency regions, experiments \cite{Fu2021quantum,Shrestha2022nontrivial,chapai2022magnetic,Broyles2022the} found effective masses typical from 0.3 to 1.7.
In general, our calculated $m^*$ at $-$40 and $-$85 meV are within experimental ranges, by considering the uncertainty of Fermi energy in experiments.
Therefore, we conclude that our calculations considering the interlayer structural modulation agree with recent quantum oscillation experiments in both fundamental frequency and cyclotron mass.

\begin{table*}
\centering
\caption{\label{Table_freq} Quantum orbits of the 2$\times$2$\times$2 CDW at $-$40 and $-$85 meV. Frequency (Freq.) is in unit of T.
$k_z$ refers to the $k_z$ plane (in unit of $2\pi/c$, $c$ is the lattice constant) the corresponding quantum orbit located in. The wavy-underlined $k_z$ indicates a minimal Fermi surface cross-section and the others correspond to maximal ones.
The maximal/minimal orbit provides helpful information to determine the sign of the extra $\pi/4$ phase shift when analyzing quantum oscillations of a three-dimensional material.
The cyclotron mass $m^*$ is in unit of bare electron mass $m_0$, where positive and negative values are for electron and hole pockets, respectively. The origin of quantum orbits (i.e., layer-contribution) is indicated by SD, ISD and mixed (SD+ISD), as obtained from Fig. \ref{Fig-band-FS}(b)-(f).
The Berry phase $\phi_B$ of the quantum orbit is calculated without SOC and the generalized Berry phase $\lambda$ is calculated with SOC (reduced to the range of --$\pi$ to $\pi$).
All quantum orbits are doubly degenerate and have $\lambda$ of the same magnitude but reversed signs (only the positive set is listed).
Depending on $|\lambda|$ being either smaller or larger than 0.5$\pi$, the final phase shift ($\Delta\phi$) in quantum oscillation contributed by two degenerate quantum orbits is either 0 or $\pi$.
The four types of non-trivial quantum orbits are defined in the main text and classified according to Table S1 and S2 in Supplementary.
}
\renewcommand\arraystretch{1.1}
\begin{ruledtabular}
\begin{tabular}{rcrccccc c rcrccccc}
  \multicolumn{8}{c}{$-$40 meV}  & & \multicolumn{8}{c}{$-$85 meV} \\
  \cline{1-8}
  \cline{10-17}
  Freq. & $k_z$ & $m^*$ & Origin  & $\phi_B$ & $\lambda$ & $\Delta\phi$ &Type& &  Freq. & $k_z$ & $m^*$ & origin & $\phi_B$ & $\lambda$ &$\Delta\phi$ &Type\\
   (T) & ($2\pi/c$) & ($m_0$) &    & ($\pi$) & ($\pi$) & ($\pi$) & & &  (T) & ($2\pi/c$) & ($m_0$) &  & ($\pi$) & ($\pi$) & ($\pi$) & \\
  \cline{1-8}
  \cline{10-17}
    13  & 0.5         & $-$0.25 & mixed  &1 &0.68 &1&I  && 5   & 0.5         &$-$0.14 & SD    &0 &0.93 &1&II \\
    20  & 0           & $-$0.43 & ISD    &0 &0.38 &0&   && 14  & \uwave{0}   &   0.21 & mixed &1 &0.77 &1&I  \\
    30  & 0.5         &    0.34 & mixed  &1 &0.83 &1&I  && 33  & \uwave{0}   &   0.25 & SD    &0 &0.84 &1&II \\
    77  & \uwave{0.5} &    0.75 & mixed  &0 &0.92 &1&III&& 40  & 0.5         &$-$0.23 & SD    &0 &0.44 &0&   \\
    98  & 0           &    0.90 & mixed  &0 &0.87 &1&IV && 51  & \uwave{0.5} &$-$0.25 & ISD   &0 &0.52 &1&IV \\
    152 & \uwave{0.5} &    0.46 & SD     &0 &0.98 &1&II && 76  & 0           &   0.27 & SD    &0 &0.02 &0&   \\
    202 & 0           &    0.66 & SD     &0 &0.37 &0&   && 99  & \uwave{0.5} &$-$0.50 & mixed &0 &0.89 &1&II \\
    230 & \uwave{0}   & $-$0.46 & mixed  &0 &0.84 &1&II && 224 & 0           &$-$0.65 & ISD   &0 &0.57 &1&III\\
    609 & \uwave{0}   & $-$0.30 & mixed  &0 &0.48 &0&   && 429 & 0.5         &   3.14 & SD    &0 &0.06 &0&   \\
    730 & 0.5         & $-$0.38 & SD     &0 &0.92 &1&II && 605 & \uwave{0}   &   0.70 & SD    &1 &0.38 &0&   \\
    734 & 0           &    1.69 & SD     &0 &0.22 &0&   && 617 & 0.5         &   0.72 & SD    &1 &0.35 &0&   \\
    894 & \uwave{0}   & $-$0.55 & mixed  &0 &0.19 &0&   && 672 & \uwave{0}   &   2.10 & mixed &0 &0.11 &0&   \\
    898 & \uwave{0.1} &    1.66 & mixed  &0 &0.71 &1&IV && 735 & \uwave{0}   &$-$0.35 & mixed &0 &0.55 &1&IV \\
    912 & 0           &    2.00 & mixed  &0 &0.31 &0&   && 765 & 0.5         &   1.09 & ISD   &0 &0.60 &1&III\\
   1067 & 0.5         &    0.90 & SD     &0 &0.33 &0&   && 940 & 0.5         &$-$0.56 & ISD   &0 &0.07 &0&   \\
   1445 & 0.5         &    0.57 & ISD    &0 &0.26 &0&   && 1267& \uwave{0}   &   0.79 & mixed &0 &0.25 &0&   \\
        &             &         &        &  &     & &   && 1293& 0.5         &   1.60 & mixed &0 &0.08 &0&   \\
\end{tabular}
\end{ruledtabular}
\end{table*}

These quantum orbits exhibit interesting interlayer hybridization, as shown in Fig. \ref{Fig-band-FS}(c)-(f) and Table \ref{Table_freq}. Approximately half of FSs at --40 and --85 meV are solely contributed by either SD or ISD layer and the rest show a strongly mixed character of two layers, labeled as $mixed$ in Table \ref{Table_freq}. There are two ways to form a mixed SD+ISD FS.
In one way, part of the Fermi contour is contributed by the SD layer and the rest by ISD with weak hybridization between two layers. This happens mostly in the $k_z = 0.5$ plane, for example, for orbits of 13 and 30 T at $-$40 meV.
The other way involves strong hybridization between two layers and the Fermi contour has comparable contributions from both layers almost everywhere.
This happens mostly in the $k_z = 0$ plane, which is the case for most mixed quantum orbits at $k_z = 0$.
Therefore, the interlayer structural modulation is crucial in determining the nature of FSs for the CsV$_3$Sb$_5$ CDW phase.
Additionally, one can find by comparing the $k_z=0$ and $0.5$ planes that many large FS pockets are quasi-2D while many small FSs are 3D in geometry.

Next, we discuss the topological properties of these quantum orbits that are debated in experiments.
In the following, the sum of the Berry phase, orbital and Zeeman phases under SOC is dubbed the generalized Berry phase $\lambda$, i.e., $\lambda = \phi_B+\phi_R+\phi_Z$. According to previous discussions \cite{Aris2018PRX}, symmetries of quantum orbits introduce additional constraints on $\lambda$. For CsV$_3$Sb$_5$ with both inversion and time-reversal symmetries, all bands are doubly degenerate with opposite generalized Berry phases $\pm \lambda$ (reduced to --$\pi \sim \pi$) in the spinful case. The superposition of the doublets leads to a $2|\mathrm{cos}(\lambda)|$ scaling of the quantum oscillation intensity, and the $\lambda$ itself is no longer the total phase shift of the quantum oscillation. Instead, the quantum oscillation always exhibits a quantized phase shift $\Delta \phi$ (see the method section and Supplementary Note 1-3). Here, $\Delta \phi$ is 0 or $\pi$, depending on whether the reduced $|\lambda|$ is smaller/larger than 0.5$\pi$.
According to the origin of $\lambda$, non-trivial quantum orbits characterized by $\Delta \phi = \pi$ (or $|\lambda|>0.5\pi$) in CsV$_3$Sb$_5$ can be classified into four types.
Type-I refers to the Dirac-point-driven quantum orbit exhibiting a robust $\phi_B=\pi$ and small $\phi_R$ and $\phi_Z$.
Type-II shows strong SOC-induced large $\phi_B$+$\phi_R$ with a small $\phi_Z$.
In type-III, the $\phi_Z$ is dominant over the $\phi_B$+$\phi_R$.
In type-IV, both the SOC-induced $\phi_B$+$\phi_R$ and $\phi_Z$ are crucial.
We point out that strong SOC and/or Zeeman effect can also suppress the $\lambda$ heavily ($|\lambda|<0.5\pi$) for some quantum orbits with a robust $\phi_B$ of $\pi$.

We first calculate the accumulated Berry phase along the quantum orbit in the 2$\times$2$\times$2 CDW phase using the Wilson loop method \cite{Fukui,Soluyanov2011,Yu2011} without including SOC. To demonstrate the type-I quantum orbits, we show the band structure without SOC in Fig.~\ref{Fig4-Dirac} (a).
At the $K/H$ point, one can find Dirac points in $k_z=0/0.5$ planes caused by the kagome structure symmetry, where we mark the higher Dirac point as $K1/H1$.
However, these Dirac points cannot be isolated band crossing points because they locate inside mirror planes in the 3D momentum space. Otherwise, they would exhibit a monopole-like Berry charge which violates the mirror symmetry. Consequently, Dirac points expand inside the mirror plane and present a continual nodal line/ring. Along the $K-H$ line, there indeed exists a Dirac nodal line connecting $K1$ and $H1$.
As shown in Fig.~\ref{Fig4-Dirac}(d), quantum orbits surrounding the nodal line show a $\pi$ Berry phase, which is the case of two medium-frequency orbits (605 and 617 T) at the Fermi energy --85 meV (Table \ref{Table_freq}).
After considering SOC, however, the generalized Berry phases $\lambda$ of these two quantum orbits are heavily reduced ($|\lambda|< 0.5\pi$) by the large Zeeman phases (roughly proportional to the cyclotron mass, see Supplementary Table S2). Thus such orbits showing no phase shifts in quantum oscillation are trivial.
Furthermore, there are additional Dirac points due to band crossings on the $M-K$ line [D$i$, $i=1\sim8$, see Fig. \ref{Fig4-Dirac}(a)] and also on the $L-H$ line (P$i$). Because of the mirror symmetry with respect to the $MKH$ plane, D$i$ and P$i$ are linked by a nodal line inside the $MKH$ plane and related bands have opposite mirror parities.
Because $k_z = 0$ and 0.5 planes are mirror planes, some of D$i$/P$i$ extend further inside the $k_z = 0/0.5$ planes and form nodal rings. Therefore, intriguing Dirac nodal networks with linked nodal lines and nodal rings emerge in the momentum space.
For example, D1 \& P1 (also P4 \& D4 and D5/6/7 \& P5/6/7) are connected to form a cylinder-like network, and P2 \& P3 are linked to constitute a 3D network.
For the nodal network, the quantum orbit can exhibit a $\pi$ Berry phase if it surrounds one nodal line and avoids crossing other nodal lines/rings. This is the case for the D5-induced FS (14 T at --85 meV) in Fig. \ref{Fig4-Dirac}(d) and P1-induced FS (30 T at --40 meV) in Fig. \ref{Fig4-Dirac}(b) which are type-I quantum orbits (Table \ref{Table_freq}).
In contrast, a Dirac-like crossing cannot generate a non-trivial FS in a plane if there is also a related nodal ring in the same plane. For example, D1/D4 fails to create a non-trivial FS in the $k_z =0$ plane. In addition, there is another nodal line crossing the $X$ point [Fig. \ref{Fig4-Dirac}(b)], leading to a $\pi$ Berry phase of the 13 T orbit at --40 meV, which belongs to the type-I quantum orbits.
These Dirac nodal lines and networks may generate non-trivial quantum orbits in a relatively large energy range besides --40 and --85 meV. We note that Dirac nodal lines were reported in the pristine phase of $A$V$_3$Sb$_5$\cite{Zhao2021,jiang2021unconventional,Hao2022dirac} rather than the CDW state.

Other three types of non-trivial quantum orbits originated from either strong SOC and/or strong Zeeman effect are shown in Table \ref{Table_freq} (see details in Supplementary Table S1\&S2). Most type-II to type-IV non-trivial quantum orbits are centered at time-reversal invariant points [i.e., $\Gamma$, $A$, $M$, and $L$ points, see Fig. \ref{Fig4-Dirac}(b)]. Thus their Berry phases $\phi_B$ and orbital phases $\phi_R$ are strictly zero without SOC due to the time-reversal symmetry constraint (see Supplementary Note 2). Only when SOC is involved the $\phi_B$ and $\phi_R$ can appear in quantum orbits.

\begin{table}
\centering
\caption{\label{Table_freq_com} Frequencies of calculated and experimentally observed non-trivial quantum orbits.
}
\renewcommand\arraystretch{1.4}
\begin{ruledtabular}
\begin{tabular}{ccccccccc}
     Exp. & $-$40 meV & $-$80 meV \\
  \hline
                                      &     & 5   \\
    18\cite{Shrestha2022nontrivial}   &  13 & 14  \\
    28\cite{Broyles2022the}           &  30 & 33  \\
    73\cite{Fu2021quantum}, 74\cite{Broyles2022the}, 79\cite{chapai2022magnetic}, 85\cite{Broyles2022the} &   77  &  51  \\
    102\cite{Shrestha2022nontrivial}  &  98 & 99  \\
                                      & 152 &     \\
    239\cite{Shrestha2022nontrivial}  & 230 & 224 \\
    727\cite{Fu2021quantum}, 736\cite{chapai2022magnetic}, 788\cite{Shrestha2022nontrivial}, 804\cite{chapai2022magnetic} & 730 &  735, 765 \\
    865\cite{Shrestha2022nontrivial}  & 898 &     \\
    1605\cite{Shrestha2022nontrivial} &     &     \\
    2135\cite{Shrestha2022nontrivial} &     &     \\
\end{tabular}
\end{ruledtabular}
\label{Table_phi}
\end{table}

Overall, we found sixteen non-trivial quantum orbits at two Fermi energies. More than half of them originate from a mixture of SD and ISD layers. We list calculated non-trivial quantum orbits and the experimental ones in Table \ref{Table_freq_com} for convenience of comparison. Except for extremely high-frequency orbits, our calculations agree well with experiments in the fundamental frequency, cyclotron mass, and topology.

\section{extended discussions}
Recently, Zhou and Wang proposed an effective single-orbital model for the $2\times2\times1$ CDW with time-reversal symmetry breaking loop-currents\cite{zhou2021chern}, where reconstructed small Chern Fermi pockets carrying concentrated Berry curvature of the Chern band emerge at $M$ points. These pockets are connected by one-quarter and three-quarters of the primitive reciprocal lattice vectors and may play a crucial role in the observed pair density wave order in the superconducting state  \cite{chen2021roton}. We verified the existence of such Fermi pockets at $M$ for the realistic $2\times2\times2$ CDW. They can be seen at --40 meV (20 T) and --85 meV (99 T) in Figs. \ref{Fig-band-FS}(g) and \ref{Fig-band-FS}(j), respectively. Moreover, we find that under SOC the 20 T pocket does not have a Berry phase without time-reversal symmetry breaking, while the 99 T pocket has a Berry phase close to $\pi$.
Upon closer examination (Supplementary Fig. S4 for FS unfolding), the 20 T pocket around $M$ at --40 meV originates from the ISD layer due to folding a pocket centered at the crossing point between $\Gamma-K$ and $M-M$ in the primitive Brillouin zone, which can be captured by a $2\times2\times1$ real CDW with ISD bond order\cite{Dong2022loop}. In contrast, the 99 T pocket in the $k_z=0.5$ plane is of mixed ISD and SD character generated by folding a pocket at the $k_z=0.25$ plane in the primitive zone, which is unique to the alternate stacking of the ISD and SD layers of the 3D CDW.

Additionally, recent Raman experiments \cite{Wu2022charge,liu2022observation} and time-resolved angle-resolved photoemission spectroscopy \cite{Doron} detected CDW-driven phonon modes. Measured CDW modes can be described by intralayer vibrations, except that the lowest-energy phonon is attributed to the interlayer Cs vibrations. Compared to phonons, electrons experience a more three-dimensional CDW structure and show stronger interlayer interaction in CsV$_3$Sb$_5$.
The present $2\times2\times2$ CDW is a minimal model to capture the interlayer coupling and  reproduce all quantum orbits measured in experiments (except large orbits due to magnetic breakdown). If we adopt the $2\times2\times 4$ model in calculations, it will bring a much larger number of quantum orbits and complicate the physics understanding.

We point out that Dirac nodal lines and networks may be a general characteristic of kagome materials. For example, the $K1-H1$ nodal line comes from the characteristic Dirac points of the kagome lattice. This scenario applies both to nonmagnetic and magnetic kagome materials. For instance, a similar $K1-H1$ nodal line was also found in the noncollinear kagome antiferromagnets Mn$_3$Sn/Ge \cite{li2022field}.
Because many mirror planes (e.g., the $MKH$ plane and $\Gamma ML$) exist, degenerate band crossing points form nodal lines inside mirror planes and nodal lines from intersecting planes may link to each other.
In the presence of strong SOC, these nodal lines are usually gapped out\cite{Burkov2011,Fang2015} and generate large Berry curvature \cite{li2020giant} or spin Berry curvature \cite{Sun2017}. Some nodal lines also lead to drumhead-like topological surface states\cite{Weng2015} on a specific facet.

\section{conclusion}

In summary, we studied the Fermi surface properties of the 3D CDW in CsV$_3$Sb$_5$. The comprehensive analysis of quantum orbits indicates that the $2\times2\times2$ CDW with interlayer structural modulation agrees broadly with recent quantum oscillation experiments in fundamental frequencies, cyclotron mass, and non-trivial phase shifts.
The non-trivial Berry phases reveal a hidden Dirac nodal network in the momentum space due to mirror symmetry protection in the spinless case.
We advanced the study of quantum oscillations in layered materials by identifying the different structural and topological origins of all quantum orbits.
Our work not only resolves the Fermi surface puzzle in CsV$_3$Sb$_5$ but also serves as the first realistic example to show the crucial role of SOC in determining the topology of quantum orbits in quantum oscillation experiments.

\section*{Methods} \label{method}

We have performed Density functional theory (DFT) calculations with the Vienna $ab$-$initio$ Simulation Package (\textsc{vasp}) \cite{VASP,PRB54p11169} and fully relaxed all structures. Except for structural relaxation, spin-orbital coupling (SOC) is considered in electronic structure calculations if otherwise stated. More details about DFT calculations are referred to Ref. \onlinecite{tan2021charge}.
High-resolution Fermi surfaces are calculated via Wannier functions \cite{wannier90} (V-$d$ and Sb-$p$ orbitals) extracted from DFT.
$k$-meshes of at least 100$\times$100$\times$100 and 100$\times$100$\times$50 are used to calculate band structures in the full BZs of 2$\times$2$\times$1 and 2$\times$2$\times$2 CDWs, respectively.
The in-plane extremal (maximal and minimal) orbits of the closed FS at different energies, which corresponds to the magnetic field parallel to the $c$ axis in quantum oscillation experiments, are determined by tracking the FS slice changes along the $k_z$ direction.
The cyclotron frequency of an electron on a closed quantum orbit is calculated according to the Onsager's relation\cite{onsager1952interpretation}, $F = \frac{\hbar c}{2\pi e} A_e$, where $A_e$ is the area of an extremal orbit at Fermi energy $\varepsilon_F$, $e$ is the electron charge and $\hbar$ is the reduced Planck constant.
The effective mass of a cyclotron electron is calculated with $m^* = \frac{\hbar^2}{2\pi m_0} \frac{\partial A_e}{\partial \varepsilon_F}$, where $m_0$ is the bare electron mass.

\textbf{Quantum oscillation phase calculation.}
Comprehensive explanations of different phases in quantum oscillations are found in Supplementary Notes 1-3. Here we provide a brief introduction on how to calculate them. More theoretical backgrounds can be found in Refs. \onlinecite{Aris2018PRX,Aris2018PRB}.

In the quantum oscillation experiment, electrons undergo cyclotron motion whose phase interference condition gives rise to Landau level. There are mainly six contributions to the total phase and the quantization rule for the total phase is,
\begin{equation}
    l_{B}^{2} A_e +\lambda_{a} - \phi_M = 2\pi n
\end{equation}
\noindent where $l_B=\sqrt{\hbar c/eB}$ is the magnetic length. The first term is the sum of the de Broglie phase and Aharonov-Bohm (A-B) phase. $\phi_M$ is the Maslov correction and equals to $\pi$ for a simple closed smooth curve (all the cyclotron orbits considered in this paper belong to this category). The generalized Berry phase $\lambda_a$ ($a = 1, 2$ the degenerate band indices) has three contributions (Berry phase $\phi_B$, orbital phase $\phi_R$ and spin Zeeman phase $\phi_Z$) and can be calculated from the eigenvalue $e^{i\lambda_a}$ of the propagator U[$C$] of loop $C$ ($\overline{\mathrm{exp}}$ means path-ordered exponential),
\begin{equation}
    U[C] = \overline{\mathrm{exp}}\left\{i\oint\left[(\boldsymbol{A} + \boldsymbol{\mathfrak{A})}\cdot d\boldsymbol{k}+\frac{g_0\hbar}{4 m v^{\perp}}\sigma^{z}|d\boldsymbol{k}|\right]\right\}
\label{Aris propagator}
\end{equation}
\noindent where $\boldsymbol{A}_{mn}=i\langle u_{mk}|\boldsymbol{\nabla}_k u_{nk}\rangle$ ($m, n\in\mathbb{Z}_D$, $\mathbb{Z}_D$ is the degenerate band group being considered.) is the non-Abelian Berry connection, and the first term is the Berry phase for the multiband case. The second phase which comes from the orbital magnetic moment is
\begin{equation}
\boldsymbol{\mathfrak{\mathfrak { A }}}_{m n} \cdot d \boldsymbol{k}=\sum_{l \notin \mathbb{Z}_D} A_{m l}^x \Pi_{l n}^y d k_x / 2 v_y+(x \leftrightarrow y)
\end{equation}
\noindent with $\boldsymbol{\Pi}_{l n} = \langle u_{lk}|\boldsymbol{v}|u_{nk}\rangle$ being the interband matrix elements of the group velocity operator $\boldsymbol{v}$. Here we emphasize the importance of the orbital phase which is usually ignored in experiments. Since CsV$_3$Sb$_5$ has inversion and time-reversal symmetries, the orbital phase is identically zero for all quantum orbits without SOC. But in the presence of SOC, it may contribute non-negligibly to the total phase $\lambda_a$. For example, it can compensate the loss of the $\pi$ Berry phase in a Dirac model with a Semenoff mass\cite{Fuchs_2010} (see also Supplementary Note 3).
The last term comes from the spin Zeeman effect and is usually written as a reduction factor\cite{shoenberg, Champel} in quantum oscillation.
When SOC is not included, the three terms can be calculated separately.
Under SOC, the three contributions to the propagator must be calculated as a whole, so separating the spin reduction factor is no longer possible. In other words, the orbital and Zeeman phases must be calculated together with the Berry phase to get observable results.

The Berry phase with or without the inclusion of the orbital phase and the generalized Berry phase $\lambda_a$ calculated for quantum orbits of the 2$\times$2$\times$2 CDW are listed in Table S1\&S2 in the Supplementary. Due to the double degeneracy of quantum orbits ensured by both inversion and time-reversal symmetries, their generalized Berry phases are $\lambda_1 = +\lambda$ and $\lambda_2 = -\lambda$. Thus the sum of two oscillations is
\begin{equation}
\begin{split}
\begin{aligned}
&\mathrm{cos}(l_B^2 A_e+\lambda_1-\phi_M \pm\frac{\pi}{4})+\mathrm{cos}(l_B^2 A_e+\lambda_2-\phi_M \pm\frac{\pi}{4})\\
=&2\mathrm{cos}(\frac{\lambda_1-\lambda_2}{2})\cdot\mathrm{cos}(l_B^2 A_e+\frac{\lambda_1+\lambda_2}{2}-\phi_M \pm\frac{\pi}{4}) \\
=&2|\mathrm{cos}(\lambda)|\cdot\mathrm{cos}(l_B^2 A_e+\Delta\phi-\phi_M \pm\frac{\pi}{4})
\end{aligned}
\end{split}
\end{equation}
where the final phase shift $\Delta\phi$ is 0 or $\pi$, depending on whether $\mathrm{cos}(\lambda)$ is positive ($|\lambda|<0.5\pi$) or negative ($|\lambda|>0.5\pi$).
If $|\lambda| = 0.5\pi$, the oscillation may not observable because $\mathrm{cos}(\lambda) = 0$.
Notice that the $\pm\frac{\pi}{4}$ phase correction comes from the $k_z$ integral in a three-dimensional case (it does not exist in a two-dimensional system).

\section*{DATA AVAILABILITY}
The data supporting the findings of this study are available within the paper and in the supplementary information.

\section*{Acknowledgement}
We thank Tobias Holder for the helpful discussions. H.T. acknowledges support from the Dean of Faculty Fellowship at Weizmann Institute of Science.
B.Y. acknowledges the financial support by the European Research Council (ERC Consolidator Grant, No. 815869).
Z.W. is supported by the U.S. Department of Energy, Basic Energy Sciences Grant No.DE-FG02-99ER45747 and a Cottrell SEED Award No. 27856 from Research Corporation for Science Advancement.

\section*{AUTHOR CONTRIBUTIONS}
B.Y. and Z.W. conceived and managed the project. The DFT and quantum oscillation calculations are calculated by H.T. The phases of quantum orbits are calculated by Y.Li. All calculations and analyses are completed with helpful input from Y.Liu and D.K. H.T., Y.Li, B.Y., and Z.W wrote the manuscript. All authors commented on the manuscript.

\section*{Competing Interests}
The Authors declare no Competing Financial or Non-Financial Interests.


%

\end{document}


\title{Supplemental Material for ``Emergent topological quantum orbits in the charge density wave phase of kagome metal CsV$_3$Sb$_5$''}

\author{Hengxin Tan}
\author{Yongkang Li}
\author{Yizhou Liu}
\author{Daniel Kaplan}
\affiliation{Department of Condensed Matter Physics, Weizmann Institute of Science, Rehovot 7610001, Israel}
\author{Ziqiang Wang}
\affiliation{Department of Physics, Boston College, Chestnut Hill, Massachusetts 02467, USA}
\author{Binghai Yan}
\affiliation{Department of Condensed Matter Physics, Weizmann Institute of Science, Rehovot 7610001, Israel}


\maketitle

\hspace*{\fill} \\

\begin{figure}
\centering
\includegraphics[width=0.65\linewidth]{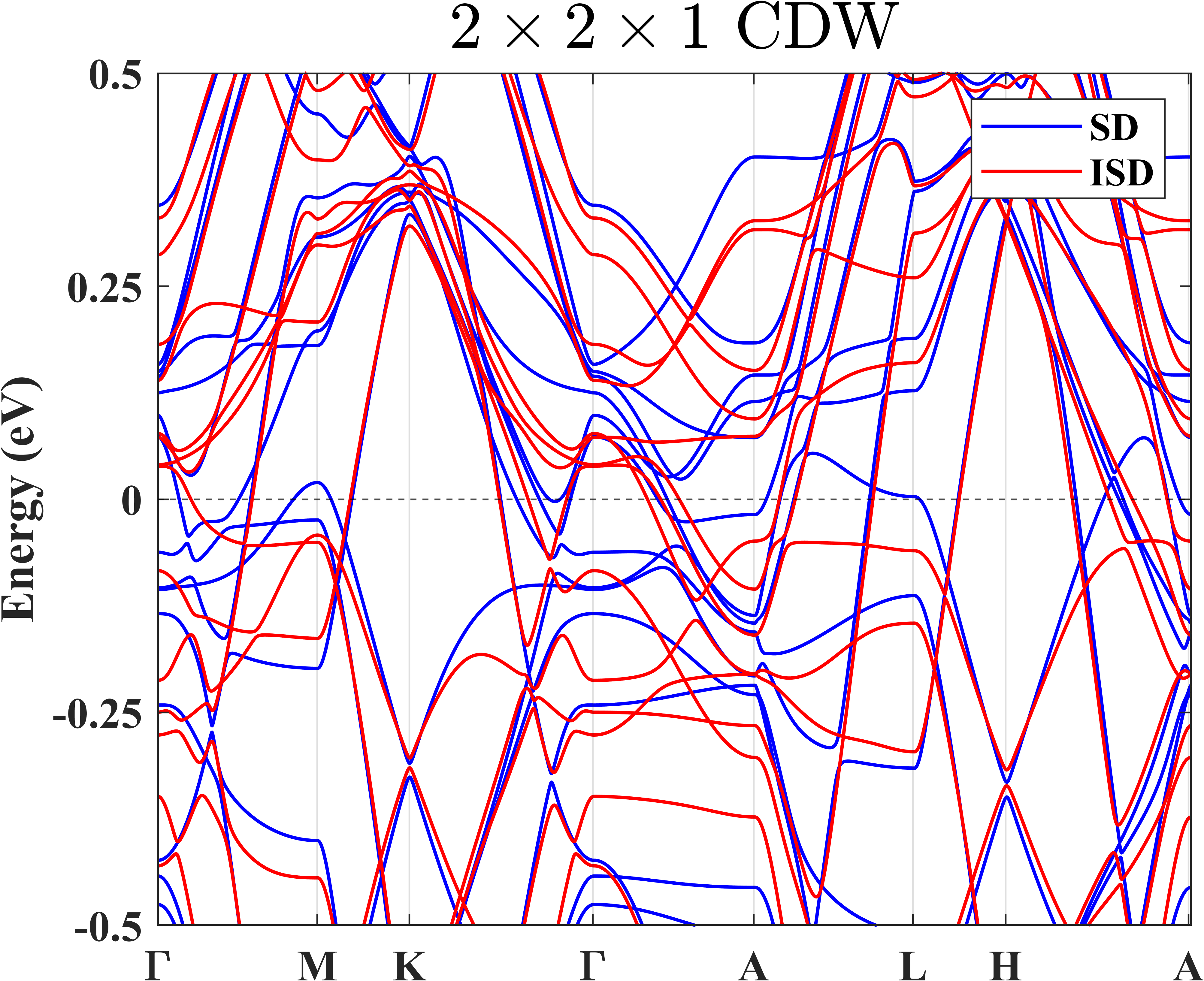}
\caption{\label{FigS1}
Overlapping band structures of 2$\times$2$\times$1 CDWs of CsV$_3$Sb$_5$, blue for SD and red for ISD. Spin orbital coupling is considered. There is no coupling between SD and ISD.
}
\end{figure}

\begin{figure}
\centering
\includegraphics[width=0.6\linewidth]{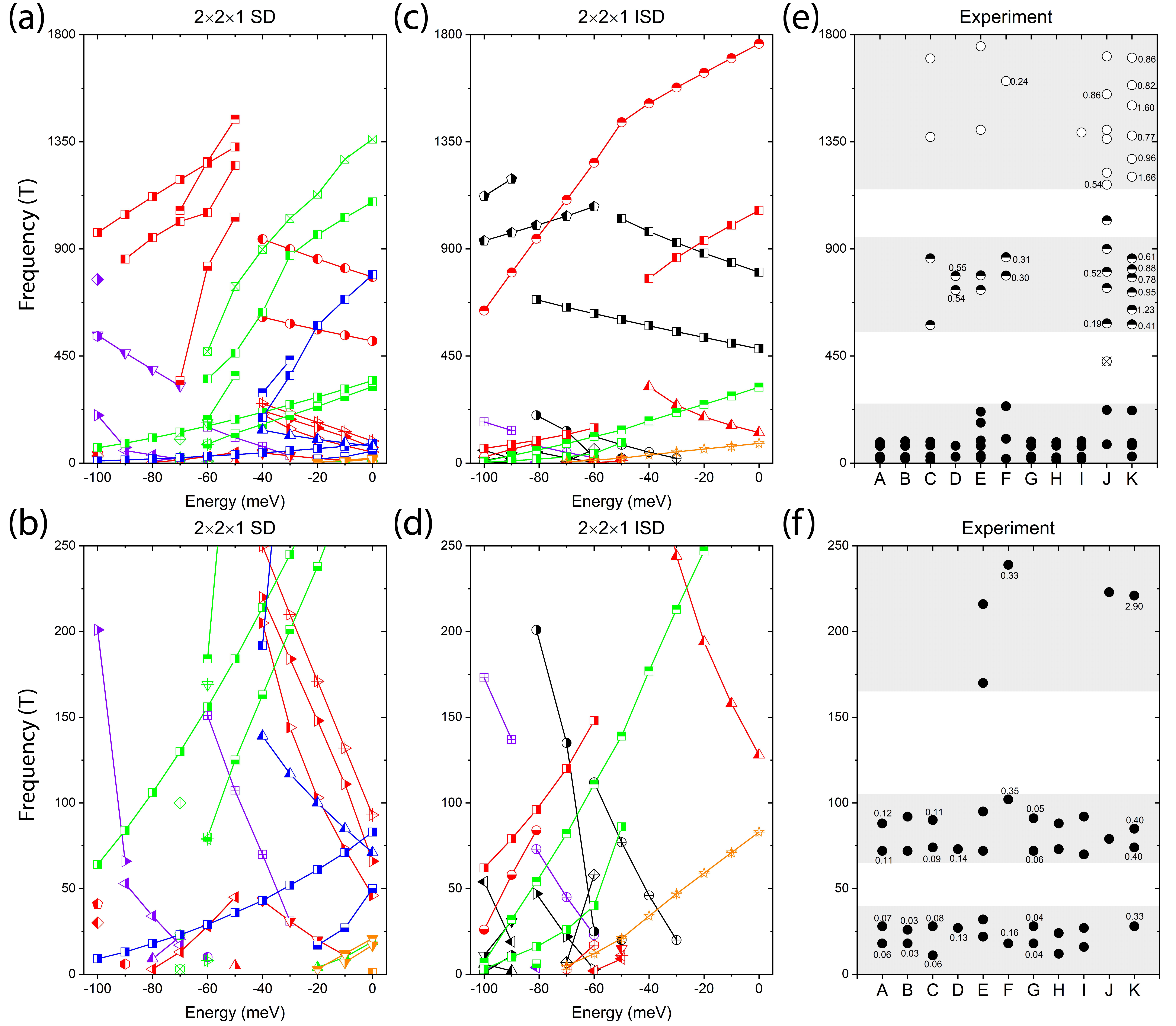}
\caption{\label{FigS2}
Cyclotron frequencies of electrons on extremal orbits of 2$\times$2$\times$1 CDWs of CsV$_3$Sb$_5$. (a) and (b) are for SD. (c) and (d) are for ISD.
(e) and (f) show the same experimental results as in the main text for comparison convenience.
(a), (c) and (e) show the full range of 0$\sim$1800 T while (b), (d), and (f) show the low-frequency region 0$\sim$250 T.
}
\end{figure}

\begin{figure}
\centering
\includegraphics[width=0.7\linewidth]{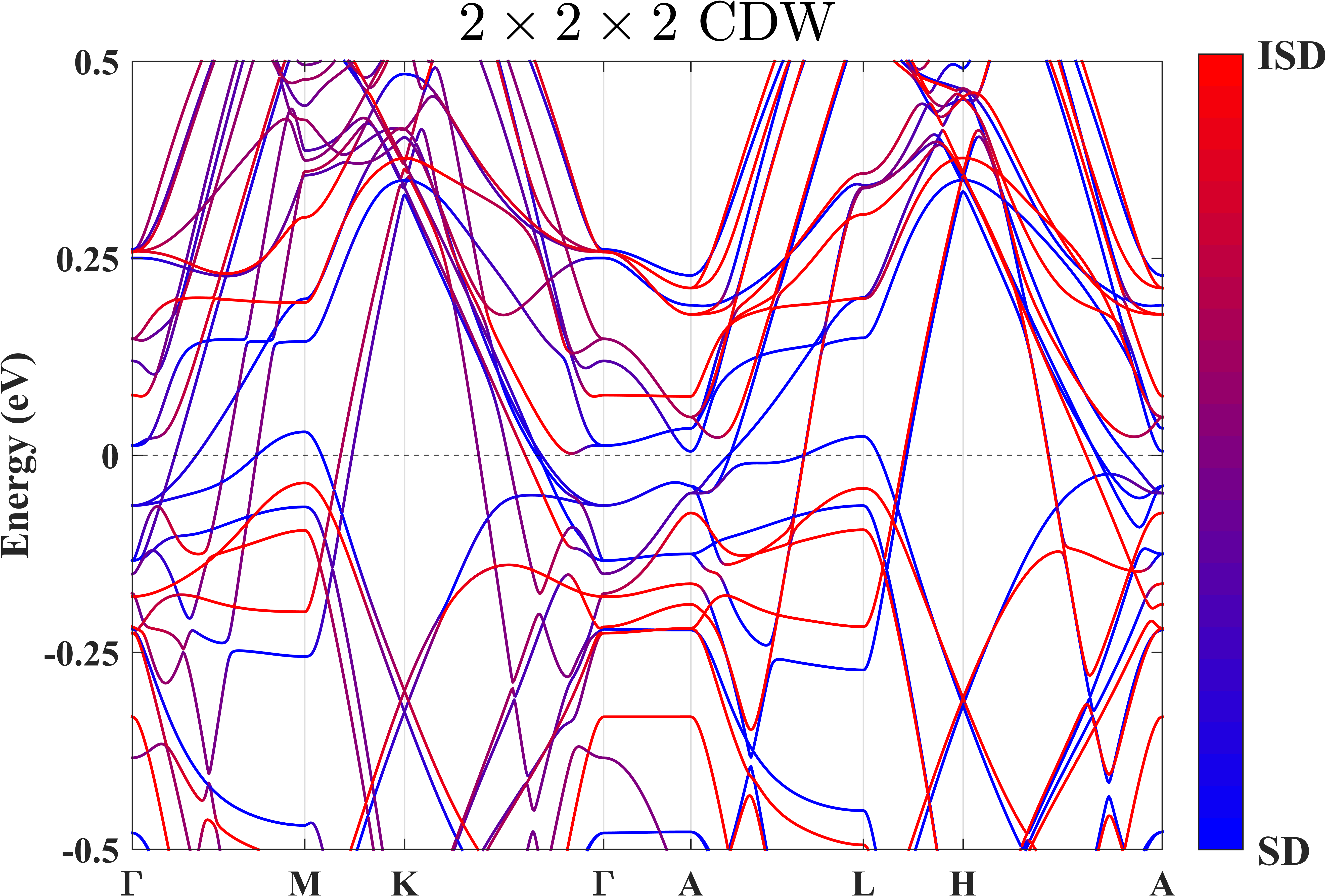}
\caption{\label{FigS3}
Layer resolved band structure of the 2$\times$2$\times$2 CDW structure of CsV$_3$Sb$_5$ without considering spin-orbital coupling.
}
\end{figure}

\begin{figure}
\centering
\includegraphics[width=0.8\linewidth]{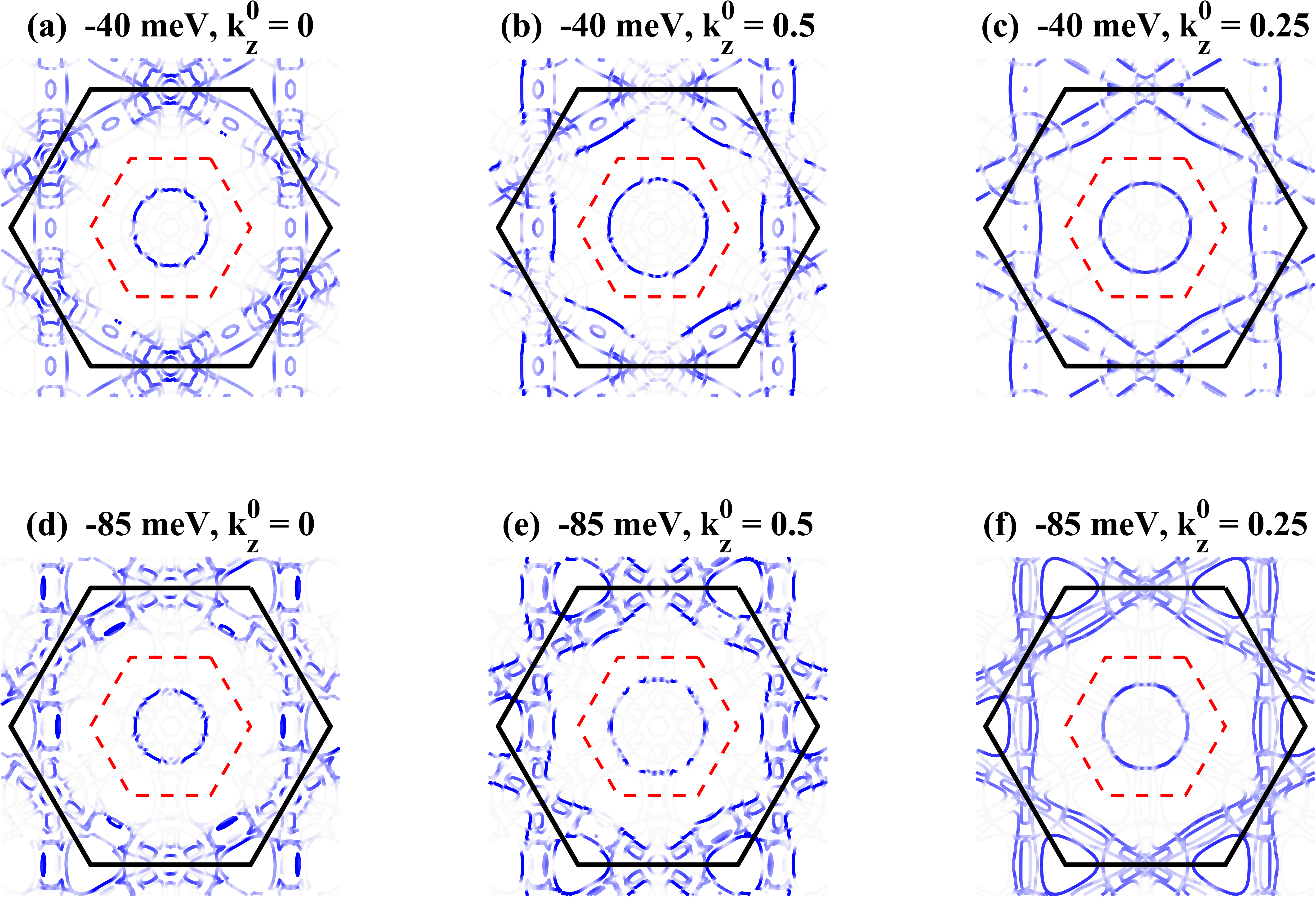}
\caption{\label{FigS4}
Unfolded Fermi surfaces of the 2$\times$2$\times$2 CDW, (a-c) for $-$40 meV and (d-f) for $-$85 meV, respectively.
The larger black solid hexagons show the Brillouin zone of the pristine phase (no CDW distortion), while the smaller red dashed hexagons show the Brillouin zone of the CDW structure.
$k_z$ and $k_z^0$ (in unit of 2$\pi$/$c$) label planes of Brillouin zones of the CDW and pristine phases, respectively.
Notice that the $k_z = 0$ plane of the 2$\times$2$\times$2 CDW phase is unfolded back to both $k_z^0 = 0$ and 0.5 planes of the pristine phase.
The $k_z = 0.5$ plane of the 2$\times$2$\times$2 CDW phase is unfolded back to the $k_z^0 = 0.25$ plane of the pristine phase.
The main features of these Fermi surfaces are similar, i.e., all show a large hexagonal Fermi surface near the Brillouin zone boundary and a circular Fermi surface around the Brillouin zone center.
}
\end{figure}

\hspace*{\fill} \\
\hspace*{\fill} \\
\textbf{Table of content for figures, notes, and tables:}
\begin{itemize}
  \item Supplementary Fig. \ref{FigS1} displays band structures of the 2$\times$2$\times$1 CDW structures SD and ISD.
  \item Supplementary Fig. \ref{FigS2} shows cyclotron frequencies of extremal orbits of 2$\times$2$\times$1 SD and ISD structures.
  \item Supplementary Fig. \ref{FigS3} displays the layer resolved band structure of the 2$\times$2$\times$2 CDW without spin-orbital coupling.
  \item Supplementary Fig. \ref{FigS4} shows the unfolded Fermi surfaces of the 2$\times$2$\times$2 CDW at $-$40 meV and $-85$ meV.
  \item Supplementary Fig. \ref{FigS5} shows all Fermi surfaces of the 2$\times$2$\times$2 CDW at different energies.
  \item Supplementary Note \ref{LK-fomula} provides a general derivation of the Lifshitz-Kosevich formula with discussions on the phase shift.
  \item Supplementary Note \ref{PT_berry} discusses the Berry phase in a time reversal and inversion symmetric system.
  \item Supplementary Note \ref{Dirac_oscillation} shows the $\pi$ total phase of a Dirac Hamiltonian with Semenoff mass.
  \item Supplementary Table \ref{Table_phi_40} and \ref{Table_phi_85} show the different phases of the quantum orbits (with and without spin-orbital coupling) of the 2$\times$2$\times$2 CsV$_3$Sb$_5$ at --40 meV and --85 meV, respectively.
\end{itemize}

\clearpage

\begin{figure}[htbp]
\centering
\subfigure{
\includegraphics[width=0.95\linewidth]{FigS5-ALL-FS-p1.png}
}
\end{figure}
\addtocounter{figure}{0}
\begin{figure}
\addtocounter{subfigure}{0}
\subfigure{
\includegraphics[width=0.95\linewidth]{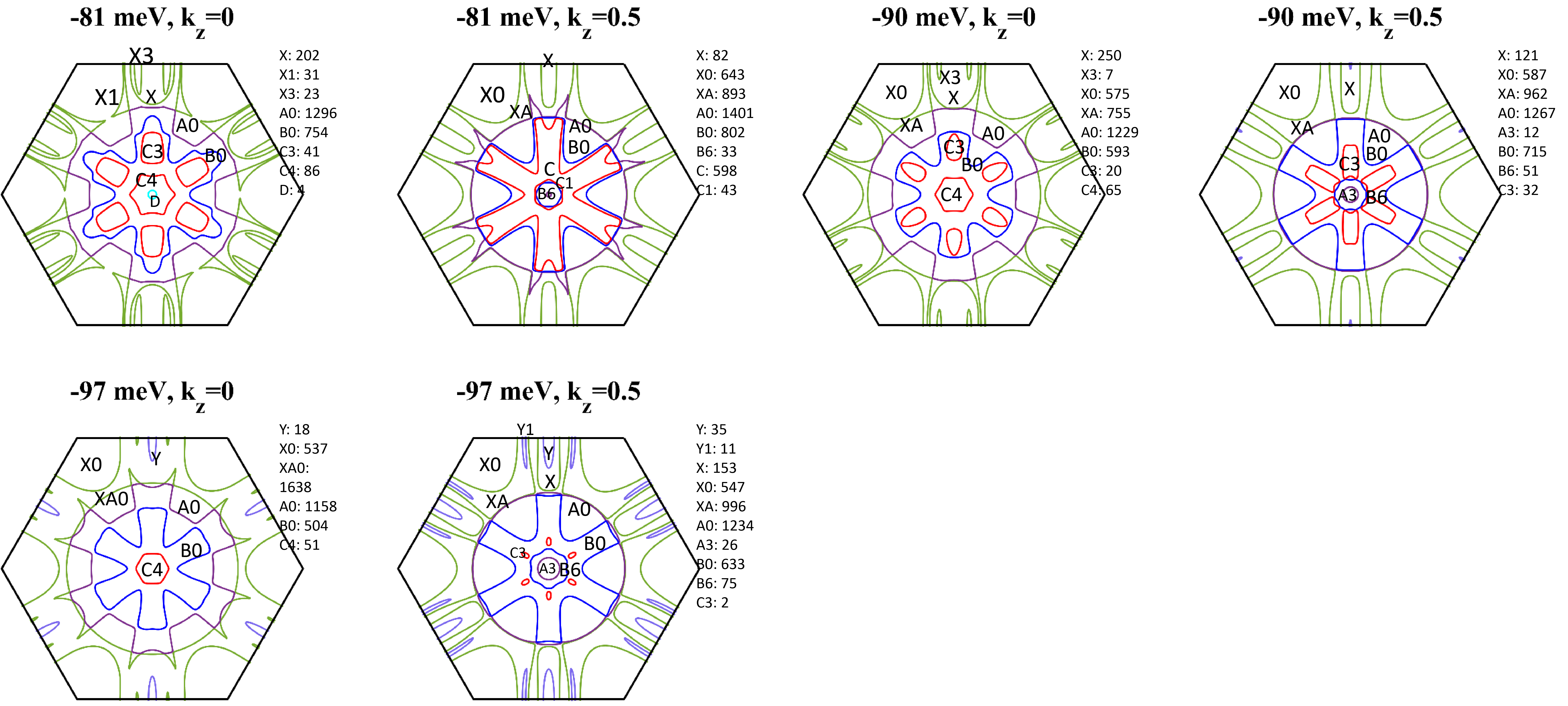}
}
\caption{\label{FigS5}
Fermi surfaces (or constant energy contours) on $k_z =$ 0 and 0.5 (2$\pi$/$c$) planes at different energies of the 2$\times$2$\times$2 CDW phase of CsV$_3$Sb$_5$ (SOC is considered). Corresponding cyclotron frequencies are given beside each panel (unit: T). Results at $-$40 and $-$85 meV are shown in the main text. Notice that frequencies smaller than 1 T are not considered.
}
\end{figure}

\clearpage

\section{Derivation of Lifshitz-Kosevich formula} \label{LK-fomula}

For convenience and completeness of reference, we briefly summarize the derivation of the Lifshitz-Kosevich formula \cite{Lifshitz} from previous works. We first give the quantization rule of Landau levels. Then an intuitive argument is provided for the final oscillatory part of the grand canonical potential. At last, a formal derivation based on Green's function method is presented.

\subsection{Semiclassical Quantization rule and magnetic oscillation}
In the presence of magnetic field $B$ (we assume it's along z direction), the electron will undergo cyclotron motion whose quantization condition gives the quantized Landau level for an arbitrary dispersion. The central requirement for the quantization of orbit is that the phase of wave function accumulated during one cycle must satisfy the interference condition. For now, the phase has mainly six contributions: de Broglie phase, Aharonov-Bohm (A-B) phase, Maslov correction $\phi_M$, Berry phase, orbital magnetic moment induced phase and Zeeman (spin) effect induced phase \cite{Fuchs_2010, Aris2018PRX,Aris2018PRB}. The de Broglie phase is $2\pi$ times the number of wavelengths in the orbit $\int{\boldsymbol{k}}\cdot{d\boldsymbol{r}}$. It's well-known that the real space trajectory of an electron is the trajectory of $k$ normal to $B$ scaled by magnetic length $l_B=\sqrt{\hbar c/eB}$ and rotated by $\SI{90}{\degree}$ \cite{Ashcroft76}. The A-B phase is the number of magnetic flux quanta $2\pi \hbar c/e$ through the trajectory multiplied by $2\pi$. For a simple trajectory, it has two caustics when projected onto one direction and each has a $\pi/2$ phase shift similar to the optical phase shift. Finally, there is an additional phase $\lambda_a$ ($a$ is the band index of degenerate bands) from the $k$ space trajectory which incorporates Berry phase $\phi_B$, orbital phase $\phi_R$ and Zeeman phase $\phi_Z$\cite{Aris2018PRX,Aris2018PRB}. We term it the generalized Berry phase. It can be calculated from the eigenvalue $e^{i\lambda_a}$ of the propagator U[$C$] of loop $C$ ($\overline{\mathrm{exp}}$ means path-ordered exponential),
\begin{equation}
    U[C] = \overline{\mathrm{exp}}\left\{i\oint\left[(\boldsymbol{A} + \boldsymbol{\mathfrak{A})}\cdot d\boldsymbol{k}+\frac{g_0\hbar}{4 m v^{\perp}}\sigma^{z}|d\boldsymbol{k}|\right]\right\}
\label{Aris propagator}
\end{equation}
\noindent where $\boldsymbol{A}_{mn}=i\langle u_{mk}|\boldsymbol{\nabla}_k u_{nk}\rangle$ ($m, n\in\mathbb{Z}_D$, $\mathbb{Z}_D$ is the degenerate band group being considered.) is the non-Abelian Berry connection and the first term is just the generalization of Berry phase to the multiband case. The second phase which comes from the orbital magnetic moment is
\begin{equation}
\boldsymbol{\mathfrak{\mathfrak { A }}}_{m n} \cdot d \boldsymbol{k}=\sum_{l \notin \mathbb{Z}_D} A_{m l}^x \Pi_{l n}^y d k_x / 2 v_y+(x \leftrightarrow y)
\label{phi_R}
\end{equation}
\noindent with $\boldsymbol{\Pi}_{l n} = \langle u_{lk}|\boldsymbol{v_k}|u_{nk}\rangle$ being the interband matrix elements of the group velocity operator $\boldsymbol{v}$. $\sigma^z$ is the spin Pauli matrix and the third phase comes from the Zeeman effect.

The overall interference condition becomes

\begin{equation}
\begin{aligned}
    \int{\boldsymbol{k}}\cdot{d\boldsymbol{r}} - 2\pi\frac{Bl_B^2A\left(\varepsilon, k_z\right)}{2\pi \hbar c /e} - \phi_M + \lambda_a = 2\pi n
\end{aligned}
\end{equation}

\noindent or

\begin{equation}
l_B^2A\left(\varepsilon_n, k_z\right)=2\pi(n+\gamma_a)
\label{EBK}
\end{equation}

\noindent where $A\left(\varepsilon, k_z\right)$ is the area of $k$ space trajectory and $\gamma_a=(\phi_M-\lambda_a) / 2 \pi$ is the index shift of Landau level. This is the Onsager relation with various phases taken into account \cite{Niu,Mikitik,Aris2018PRX,Aris2018PRB}.

Following the argument of Ref. \onlinecite{Ashcroft76}, equation \eqref{EBK} specifies a set of Landau tubes (for each $n$) on which quantized orbits lie. The contribution to the density of states (DOS) $D(\varepsilon_F)$ at the Fermi level $\varepsilon_F$ (at zero temperature) from these Landau levels are the number of quantized orbits with energy in a small shell near $\varepsilon_F$. Therefore, $D(\varepsilon_F)$ is proportional to the intersection area between the tubes and the small energy shell around the Fermi surface. When the tube is tangential to the Fermi surface, the area will be greatly enhanced and $D(\varepsilon_F)$ will be singular, i.e.

\begin{equation}
A_e\left(\varepsilon_F\right)=(n+\gamma_a) \frac{2\pi e B}{\hbar c}
\label{Onsager}
\end{equation}

\noindent where $A_e\left(\varepsilon_F\right)$ is the area of extremal orbit on the Fermi surface. As a consequence, the oscillation of $D(\varepsilon_F)$ is periodic in $\frac{1}{B}$ with period $\Delta\left(\frac{1}{B}\right)=2 \pi e/(\hbar c A_e\left(\varepsilon_F\right))$ and frequency

\begin{equation}
F=1/\Delta\left(\frac{1}{B}\right)=\frac{\hbar c}{2\pi e}A_e\left(\varepsilon_F\right)
\label{F}
\end{equation}

\subsection{Intuitive argument}

The oscillation of various thermal quantities such as magnetic susceptibility follows the oscillation of $D(\varepsilon_F)$, or equivalently the grand canonical potential. In general, such an oscillation part $\delta\Omega(T=0,\varepsilon_F)$ can be expanded as a Fourier series with the positions of maximum determined from \eqref{Onsager} and fundamental frequency given by \eqref{F}, i.e.

\begin{equation}
\delta\Omega(T=0,\varepsilon_F) =\sum_{r} \Omega_{r}(T=0,\varepsilon_F) = \sum_{r} c_r \ \mathrm{cos}[2\pi r (\frac{F}{B} - \gamma_a) + \varphi] = \sum_{r} c_r \ \mathrm{cos}[r (l_B^2 A_e\left(\varepsilon_F\right) + \lambda_a - \phi_M)+ \varphi]
\label{omega}
\end{equation}

\noindent where $c_r$ is the amplitude of $r$'th harmonics and $\varphi$ is an additional phase shift for for three-dimensional (3D) materials (see below).

The effect of finite temperature and finite relaxation time can be taken into account by the method of phase smearing \cite{shoenberg} and contribute a reduction factor for the oscillation amplitude. Here we adopt a different way to get the reduction factor. First, the grand canonical potential at finite T can be obtained from the zero temperature one by convolution with the derivative of the Fermi distribution \cite{shoenberg}.

\begin{equation}
\delta\Omega(T, \mu)=-\int_{-\infty}^{\infty} \frac{\partial f(\varepsilon-\mu)}{\partial \varepsilon} \delta\Omega(0, \varepsilon) d \varepsilon
\end{equation}

\noindent where $f(\varepsilon-\mu)=1/[\mathrm{exp}((\varepsilon-\mu)/k T)+1]$ is Fermi-Dirac distribution function and $\mu$ is the chemical potential at finite temperature. Substitute \eqref{omega} into the above formula and make use of the relation \eqref{F} to change integral variable from $\varepsilon$ to $F$, one can get
\begin{equation}
    \begin{aligned}
    \Omega_{r}(T,\mu)&=-c_r \mathrm{Re} \int_{-\infty}^{\infty} \frac{\partial f(\varepsilon-\mu)}{\partial \varepsilon} \mathrm{exp}[i(2\pi r(\frac{F}{B} - \gamma_a) + \varphi)] d \varepsilon \\
    &=c_r \mathrm{Re} \int_{-\infty}^{\infty} \frac{1}{2k_B T[1+\mathrm{cosh}((\varepsilon-\mu)/k_B T)]} \mathrm{exp}[i(2\pi r(\frac{F}{B} - \gamma_a) + \varphi)] d \varepsilon \\
    &=c_r \mathrm{Re} \int_{-\infty}^{\infty} \frac{1}{2[1+\mathrm{cosh}((\varepsilon-\mu)/k_B T)]} \mathrm{exp}[i(2\pi r(\frac{F}{B} - \gamma_a) + \varphi)] \frac{2\pi e}{k_B T \hbar c A^{'}_{e}(\mu)} d F \\
    &=c_r \mathrm{Re} \int_{-\infty}^{\infty} \frac{1}{1+\mathrm{cosh}[\beta(F-F(\mu))/k_B T]} \mathrm{exp}[i(2\pi r(\frac{F}{B} - \gamma_a) + \varphi)] \frac{\beta}{2k_B T} d F \\
    &=c_r \mathrm{Re}\int_{-\infty}^{\infty} \mathrm{exp}[i(2\pi r(\frac{F(\mu)}{B} - \gamma_a) + \varphi)] \cdot \frac{\mathrm{exp}[i 2\pi r k_B Tz/(\beta B)]}{2[1+cosh(z)]} d z\\
    &=c_r R_{T,r} \, \mathrm{cos}[2\pi r(\frac{F(\mu)}{B} - \gamma_a) + \varphi]\\
    &=R_{T,r} \, \Omega_{r}(0,\mu)
    \end{aligned}
\end{equation}

\noindent with

\begin{equation}
R_{T,r}=\frac{2 \pi^2 r k_B T / \beta B}{\sinh \left(2 \pi^2 r k_B T / \beta B\right)}, \quad \beta = \frac{e\hbar}{m c}, \quad m=\frac{\hbar^2}{2\pi}\frac{\partial A_e(\varepsilon, k_z)}{\partial \varepsilon}
\label{mc}
\end{equation}

\noindent $m$ is the cyclotron mass. Use cyclotron frequency $\omega_c=eB/mc$, temperature reduction factor $R_{T,r}$ can also be written as

\begin{equation}
R_{T,r}=\frac{2 \pi^2 r k_B T / \hbar \omega_c}{\sinh \left(2 \pi^2 r k_B T / \hbar \omega_c\right)}
\label{RT}
\end{equation}

Second, a finite relaxation time can be qualitatively considered as the damping of wave function amplitude. The grand canonical potential is proportional to the DOS, which is further proportional to the diagonal elements of Green's function $G(\boldsymbol{r},\boldsymbol{r},E)$. The latter one represents the returning probability amplitude for electron starting from $\boldsymbol{r}$ and returning to $\boldsymbol{r}$. If We assume the wave packets decay in time like \cite{Falicov}

\begin{equation}
    |\Psi(t)| = \mathrm{exp}(-t/2\tau)
\label{tau}
\end{equation}

\noindent then the oscillation amplitude is reduced by the same factor with $t$ the time for an electron to return to the original position. For $r$'th harmonics, the time for returning is the time for completing $r$ cycles: $t_r = 2\pi r/ \omega_c$ ($\omega_c$ is cyclotron frequency). Thus the relaxation time reduction factor (Dingle factor) is \cite{Falicov}

\begin{equation}
    R_{D,r} = \mathrm{exp}(-t_r/2\tau) = \mathrm{exp}(-\pi r / \omega_c \tau)
\label{RD}
\end{equation}





Put all these relations together, we get the generalized Lifshitz-Kosevich formula (for one band)\cite{Aris2018PRX}

\begin{equation}
\begin{aligned}
    \delta\Omega(T,\mu) = \sum_{r} c_r \ \frac{2 \pi^2 r k_B T / \hbar\omega_c}{\sinh \left(2 \pi^2 l k_B T / \hbar\omega_c\right)} \ \mathrm{exp}(-\frac{\pi r}{\omega_c \tau}) \ \mathrm{cos}[2\pi r (\frac{F}{B} - \gamma_a) + \varphi]
\end{aligned}
\label{omega1}
\end{equation}

\subsection{Formal derivation}
For a formal justification of formula \eqref{omega1} and also the determination of the amplitude $c_r$, we provide a direct calculation of the grand canonical potential taking the finite T and $\tau$ into account at the beginning. Here the method is from Champel and Mineev \cite{Champel}. First, we focus on one band taking possible spin-orbit coupling into consideration (so the wave function is a two-component spinor). In the presence of disorder scattering, if one uses the Born approximation, then the self-energy is

\begin{equation}
    \Sigma^{R}(\varepsilon, k_z)= - i\Gamma= - i\frac{\hbar}{2\tau}
\end{equation}

\noindent where the real part of self-energy is ignored whose effect is only a constant shift of energy in this approximation. $\tau$ is the relaxation time used in \eqref{tau} and gives the Dingle factor as can be seen later. The retarded Green's function is now modified by the self-energy $\Sigma^{R}(\varepsilon, k_z)$ as

\begin{equation}
    G_n^{R}(\varepsilon, k_z)=\frac{1}{\varepsilon - \varepsilon_n(k_z) - \Sigma^{R}(\varepsilon, k_z)}
\end{equation}

\noindent where the Landau level $\varepsilon_n(k_z)$ is determined by \eqref{EBK}. The spectral density is the imaginary part of Green's function

\begin{equation}
\begin{aligned}
    S_n(\varepsilon, k_z)&=-\frac{1}{\pi} \mathrm{Im} G_n^R(\varepsilon, k_z) \\
    &=\frac{1}{\pi} \frac{\Gamma}{[\varepsilon - \varepsilon_n(k_z)]^2+\Gamma^2}
\end{aligned}
\end{equation}

The spectral density function can be used to define a generalized density of states $D(\varepsilon)$ whose integral is the particle number. In the present case, each Landau level has a degeneracy $D=e B/2\pi \hbar c$, and $D(\varepsilon)$ is given by

\begin{equation}
\begin{aligned}
    D(\varepsilon)=D\int \frac{d k_z}{2\pi}\sum_{n} S_n(\varepsilon, k_z)
\end{aligned}
\end{equation}

To calculate the sum over $n$, we first define a continuous variable $x$ via

\begin{equation}
A\left(\varepsilon(x), k_z\right)=x \frac{2\pi e B}{\hbar c}
\label{Onsagerx}
\end{equation}

\noindent i.e. $n+\gamma_a$ in \eqref{EBK} is replaced by $x$. Then with the help of Poisson summation formula\cite{shoenberg} ($0\leq\gamma_a\leq 1$)

\begin{equation}
    \sum_{n=0}^{\infty} f(n+\gamma_a)=\int_{0}^{+\infty}f(x) d x + 2\sum_{r=1}^{\infty}\int_{0}^{+\infty}f(x) \mathrm{cos}\left[2\pi r(x-\gamma_a)\right] d x
\end{equation}

\noindent $D(\varepsilon)$ can be written as

\begin{equation}
\begin{aligned}
    D(\varepsilon)&=\frac{D}{\pi}\int \frac{d k_z}{2\pi}\left[ \int_{0}^{+\infty}d x \frac{\Gamma}{[\varepsilon - \varepsilon(x, k_z)]^2+\Gamma^2} + 2 \mathrm{Re} \sum_{r=1}^{\infty}\int_{0}^{+\infty}d x \frac{\Gamma}{[\varepsilon - \varepsilon(x,k_z)]^2+\Gamma^2} \mathrm{exp}(2\pi i r(x-\gamma_a)) \right]
\end{aligned}
\label{DOS}
\end{equation}

The first part of the above formula represents the DOS in the absence of a magnetic field, and the second part is the oscillation part which we shall focus on. Below we calculate the $r$'th Fourier components in the oscillation part $D_r(\varepsilon)$. Changing the variable from $x$ to $\varepsilon(x,k_z)$ via \eqref{Onsagerx} and use the definition of effective mass $m(\varepsilon)$ \eqref{mc}, the integral is written as

\begin{equation}
    D_r(\varepsilon)=\frac{1}{\pi^2 \hbar^2} \mathrm{Re} \int \frac{d k_z}{2\pi} \int_{\varepsilon_0}^{+\infty}d \varepsilon^{'} \frac{\Gamma}{[\varepsilon - \varepsilon^{'}]^2+\Gamma^2} m(\varepsilon^{'}) \, \mathrm{exp}\left[2\pi i r(\frac{\hbar c}{2\pi e B}A(\varepsilon^{'}, k_z)-\gamma_a) \right]
\label{DOSr}
\end{equation}

\noindent The lower integral limit $\varepsilon_0$ is the energy for the cross section area to vanish, i.e., $A\left(\varepsilon_0, k_z\right)=0$. In general, $\varepsilon_0$ depends on $k_z$, but if we assume $\Gamma << \varepsilon$, then the contribution of the integral over $\varepsilon^{'}$ is mainly from the vicinity of $\varepsilon$, hence the lower integral limit can be extended to $-\infty$. Then interchange the order of integral and do the integral over $k_z$ first, we need to calculate

\begin{equation}
    \int \frac{d k_z}{2\pi} \mathrm{exp}\left[2\pi i r(\frac{\hbar c}{2\pi e B}A(\varepsilon^{'}, k_z)-\gamma_a) \right]
\end{equation}

This integral can be done by using the stationary phase approximation. Physically, the integral is a superposition of oscillation of DOS from all different $k_z$. If the cross-section area $A(\varepsilon^{'}, k_z)$ changes rapidly, then these contributions cancel out. The only exception is the extremal orbit where $d A(\varepsilon^{'}, k_z)/d k_z$ vanishes and the oscillation from the orbits around it adds coherently. Hence, after integrating over $k_z$, only the extremal orbit has an appreciable contribution. The result is simply

\begin{equation}
    \sqrt{\frac{e B}{2\pi\hbar c r|A_e^{''}|}} \mathrm{exp}\left[2\pi i r(\frac{\hbar c}{2\pi e B}A_e(\varepsilon^{'})-\gamma_a)\pm i \frac{\pi}{4} \right]
\end{equation}

\noindent where $A_e(\varepsilon^{'})$ is the area of extremal orbit and there is an additional phase shift $\pm \frac{\pi}{4}$ ($+$ for minimum area and $-$ for maximum area). Substitute this result into \eqref{DOSr}, and complete the integral, the $r$'th component is found as

\begin{equation}
\begin{aligned}
    D_r(\varepsilon)&=\frac{1}{\pi^2 \hbar^2}\sqrt{\frac{e B}{2\pi\hbar c r|A_e^{''}|}} \mathrm{Re} \int_{-\infty}^{+\infty}d \varepsilon^{'} \frac{\Gamma}{[\varepsilon - \varepsilon^{'}]^2+\Gamma^2} m(\varepsilon^{'})\mathrm{exp}\left[2\pi i r(\frac{\hbar c}{2\pi e B}A_e(\varepsilon^{'})-\gamma_a)\pm i \frac{\pi}{4} \right]\\
    &\approx \frac{m(\varepsilon)}{\pi^2 \hbar^2}\sqrt{\frac{e B}{2\pi\hbar c r|A_e^{''}|}} \mathrm{Re} \ \mathrm{exp}\left[2\pi i r(\frac{\hbar c}{2\pi e B}A_e(\varepsilon)-\gamma_a)\pm i \frac{\pi}{4} \right] \int_{-\infty}^{+\infty}d \varepsilon^{'} \frac{\Gamma}{[\varepsilon^{'} - \varepsilon]^2+\Gamma^2}\mathrm{exp}\left[i \frac{r\hbar c}{e B}A_e^{'}(\varepsilon)(\varepsilon^{'}-\varepsilon) \right]\\
    &= \frac{m(\varepsilon)}{\pi^2 \hbar^2}\sqrt{\frac{e B}{2\pi\hbar c r|A_e^{''}|}} \mathrm{Re} \ \mathrm{exp}\left[2\pi i r(\frac{\hbar c}{2\pi e B}A_e(\varepsilon)-\gamma_a)\pm i \frac{\pi}{4} \right] \int_{-\infty}^{+\infty}d y \frac{\Gamma}{y^2+\Gamma^2}\mathrm{exp}\left[i \frac{2\pi r}{\hbar\omega_c}y \right] \\
    &= \frac{m(\varepsilon)}{\pi^2 \hbar^2}\sqrt{\frac{e B}{2\pi\hbar c r|A_e^{''}|}} \mathrm{Re} \ \mathrm{exp}\left[2\pi i r(\frac{\hbar c}{2\pi e B}A_e(\varepsilon)-\gamma_a)\pm i \frac{\pi}{4} \right] 2\pi i \cdot \frac{\Gamma\mathrm{exp}(-\frac{2\pi r\Gamma}{\hbar\omega_c})}{2i\Gamma}\\
    &=(\frac{e B}{2\hbar c})^{1/2} \frac{m(\varepsilon)}{\pi^{3/2}\hbar^2\sqrt{r|A_e^{''}|}} \mathrm{cos}\left[2\pi r(\frac{F}{B}-\gamma_a) \pm\frac{\pi}{4} \right] \mathrm{exp}(-\frac{\pi r}{\omega_c \tau})
\end{aligned}
\end{equation}

\noindent where $\omega_c$ is cyclotron frequency of the extremal orbit with energy $\varepsilon$ and $F=\hbar c A_e(\varepsilon)/2\pi e$. The last exponential factor is just the Dingle factor \eqref{RD}.

To proceed, the particle number at chemical potential $\mu$ (at finite temperature) is $N(\mu) = \int D(\varepsilon) f(\varepsilon-\mu) d \varepsilon$, and use thermal dynamic relation $N = -\frac{\partial \Omega(T, \mu)}{\partial \mu}|_T$. The $r$'th component of grand canonical potential (per unit volume) is

\begin{equation}
    \begin{aligned}
    \frac{\Omega_r(T,\mu)}{V}&=-\int_{-\infty}^{\mu}d \zeta \int_{-\infty}^{+\infty} d \varepsilon D_r(\varepsilon) f(\varepsilon - \zeta) \\
    &=-k_B T \int_{-\infty}^{+\infty} d \varepsilon D_r(\varepsilon) \ \mathrm{ln}\left\{1 + \mathrm{exp}\left[-(\varepsilon-\mu)/k_B T\right]\right\} \\
    &=-(\frac{e B}{2\hbar c})^{1/2} \frac{m}{\pi^{3/2}\hbar^2\sqrt{r|A_e^{''}|}} \mathrm{exp}(-\frac{\pi r}{\omega_c \tau}) k_B T \, \mathrm{Re} \int_{-\infty}^{+\infty} d \varepsilon \ \mathrm{exp}\left[2\pi i r(\frac{F(\varepsilon)}{B}-\gamma_a) \pm i\frac{\pi}{4} \right]  \mathrm{ln}\left\{1 + \mathrm{exp}\left[-(\varepsilon-\mu)/k_B T\right]\right\}
    \end{aligned}
\end{equation}

\noindent In the above formula, all the prefactors are evaluated at $\mu$ since most contribution comes from the Fermi surface. The integral is done by integration by parts twice and evaluate $F'(\varepsilon)$ at chemical potential $F'(\mu)=\beta^{-1}=B/\hbar\omega_c$ and expand $F(\varepsilon)$ around $\mu$.

\begin{equation}
    \begin{aligned}
    &k_B T\int_{-\infty}^{+\infty} d \varepsilon \ \mathrm{exp}\left[2\pi i r(\frac{F(\varepsilon)}{B}-\gamma_a) \pm i\frac{\pi}{4}) \right]  \mathrm{ln}\left\{1 + \mathrm{exp}\left[-(\varepsilon-\mu)/k_B T\right]\right\} \\
    =&k_B T(\frac{\hbar\omega_c}{2\pi i r})^2 (-\frac{1}{k_B T})^2 \frac{1}{2} \int_{-\infty}^{+\infty} d \varepsilon \frac{\mathrm{exp}\left[2\pi i r(\frac{F(\varepsilon)}{B}-\gamma_a) \pm i\frac{\pi}{4} \right]}{1+\mathrm{cosh}\left[(\varepsilon-\mu)/k_B T\right]}\\
    =&-(\frac{\hbar\omega_c}{2\pi r})^2 \mathrm{exp}\left[2\pi i r(\frac{F(\mu)}{B}-\gamma_a) \pm i\frac{\pi}{4} \right] \frac{1}{2 k_B T} \int_{-\infty}^{+\infty} d \varepsilon \frac{\mathrm{exp}\left[2\pi i r(\varepsilon-\mu)/\hbar\omega_c
    \right]}{1+\mathrm{cosh}\left[(\varepsilon-\mu)/k_B T\right]} \\
    =&-(\frac{\hbar\omega_c}{2\pi r})^2 \mathrm{exp}\left[2\pi i r(\frac{F(\mu)}{B}-\gamma_a) \pm i\frac{\pi}{4} \right]R_{T,r}
    \end{aligned}
\end{equation}

Hence,

\begin{equation}
    \begin{aligned}
    \frac{\Omega_r(T,\mu)}{V}=\frac{\hbar \omega_c}{(2\pi l_B^2)^{3/2}} \frac{1}{2\pi^2 r^{5/2}\sqrt{|A_e^{''}|}} \cdot \mathrm{cos}\left[2\pi r(\frac{F(\mu)}{B}-\gamma_a) \pm \frac{\pi}{4} \right] R_{T,r} R_{D,r}
    \end{aligned}
\label{omega2}
\end{equation}

This is the same as \eqref{omega1} and gives the explicit expression for amplitude $c_r$ and phase shift $\varphi$, which verifies our previous intuitive argument 3D system. The oscillatory part of total grand canonical potential $\delta\Omega(T,\mu)$ is the sum of $\Omega_r$ over Fourier components $r$ and degenerated bands $a$. The magnetization can be obtained by differentiating with B in the cosine factor (this is the dominant part): $M=-1/V (\partial \Omega/\partial B)$.

\begin{equation}
\begin{aligned}
    \frac{\delta\Omega(T,\mu)}{V} &= \sum_{a}\sum_{r} \frac{\hbar \omega_c}{(2\pi l_B^2)^{3/2}} \frac{1}{2\pi^2 r^{5/2}\sqrt{|A_e^{''}|}} \mathrm{cos}\left[2\pi r(\frac{F}{B}-\gamma_a) \pm \frac{\pi}{4} \right] R_{T,r} R_{D,r}\\
    &= \sum_{a}\sum_{r} \frac{\hbar \omega_c}{(2\pi l_B^2)^{3/2}} \frac{1}{2\pi^2 r^{5/2}\sqrt{|A_e^{''}|}} \mathrm{cos}\left[r(l_B^2 A_e+\lambda_a-\phi_M) \pm \frac{\pi}{4} \right] R_{T,r} R_{D,r} 
\end{aligned}
\label{omega3D}
\end{equation}

\begin{equation}
\begin{aligned}
    M(T,\mu) &= -\sum_{a}\sum_{r} \frac{\hbar \omega_c}{(2\pi l_B^2)^{3/2}} \frac{F}{\pi r^{3/2}B^2\sqrt{|A_e^{''}|}} \mathrm{sin}\left[2\pi r(\frac{F}{B}-\gamma_a) \pm \frac{\pi}{4} \right] R_{T,r} R_{D,r} \\
    &= -\sum_{a}\sum_{r} \frac{\hbar \omega_c}{(2\pi l_B^2)^{3/2}} \frac{F}{\pi r^{3/2}B^2\sqrt{|A_e^{''}|}} \mathrm{sin}\left[r(l_B^2 A_e+\lambda_a-\phi_M) \pm \frac{\pi}{4} \right] R_{T,r} R_{D,r} 
\end{aligned}
\label{M3D}
\end{equation}

\noindent $F$ takes value at $\mu$ and $R_{T,r}, R_{D,r}$ are given by \eqref{RT}, \eqref{RD} respectively.

If the spin-orbital coupling (SOC) is very weak or zero, each band is degenerate for two spin channels (one band with two electrons). There is only one Berry phase $\phi_B$, one orbital phase $\phi_R$ and two opposite Zeeman phases $\phi_Z$. Then the generalized Berry phase $\lambda_a$ can be factorized into two parts: $\tilde{\lambda}=\phi_B+\phi_R$ and $\pm \phi_Z$. The total grand canonical potential of one band is

\begin{equation}
\begin{aligned}
    \frac{\delta\tilde{\Omega}(T,\mu)}{V} &=\sum_{r} \frac{\hbar \omega_c}{(2\pi l_B^2)^{3/2}} \frac{1}{2\pi^2 r^{5/2}\sqrt{|A_e^{''}|}} \left\{\mathrm{cos}\left[r(l_B^2 A_e+\tilde{\lambda}+\phi_Z-\phi_M) \pm \frac{\pi}{4} \right] + \mathrm{cos}\left[r(l_B^2 A_e+\tilde{\lambda}-\phi_Z-\phi_M) \pm \frac{\pi}{4} \right]\right\} R_{T,r} R_{D,r}\\
    &= \sum_{r} \frac{\hbar \omega_c}{(2 \pi l_B^2)^{3/2}} \frac{1}{\pi^2 r^{5/2}\sqrt{|A_e^{''}|}} \mathrm{cos}\left[r(l_B^2 A_e+\tilde{\lambda}-\phi_M) \pm \frac{\pi}{4} \right] R_{T,r} R_{D,r} R_{s,r} 
\end{aligned}
\label{omega3D-noSOC}
\end{equation}

\noindent where

\begin{equation}
    R_{s,r}=\mathrm{cos}(r\phi_Z)=\mathrm{cos}(\frac{r\pi g m}{2m_0})
\label{Rs}
\end{equation}

\noindent is just the spin reduction factor\cite{Lifshitz, shoenberg}. One important difference between $R_s$ and $R_T$, $R_D$ is that the latter is always positive while the former can be negative. The negativity of $R_s$ for the first harmonic means an additional $\pi$ phase shift in the oscillation peaks and is usually considered separately to extract the phase $\tilde{\lambda}$.

If SOC is strong, the wave function for each band is a two-component spinor, so the separation of a spin reduction factor is no longer possible. Further, if all bands are degenerate, $\lambda_a$ must be calculated as a whole to ensure gauge invariant results. In the case of spin degeneracy ($a=1,2$), the sum of two fundamental harmonics is still a harmonic with the same frequency but reduced amplitude

\begin{equation}
    \mathrm{cos}\left[r(l_B^2 A_e+\lambda_1-\phi_M) \pm \frac{\pi}{4}\right]+\mathrm{cos}\left[r(l_B^2 A_e+\lambda_2-\phi_M) \pm \frac{\pi}{4}\right]=2|\mathrm{cos}(\frac{\lambda_1-\lambda_2}{2})|\cdot\mathrm{cos}\left[r(l_B^2 A_e+\Delta \phi-\phi_M) \pm \frac{\pi}{4}\right]
\label{spinsum}
\end{equation}

\noindent where

\begin{equation}
    \Delta \phi=\frac{\lambda_1+\lambda_2}{2}+\pi\left\{1-\eta\left[\mathrm{cos}(\frac{\lambda_1-\lambda_2}{2})\right]\right\}
\label{spinsum}
\end{equation}

\noindent with $\eta(x)$ the Heaviside step function. In the general case, $|\mathrm{cos}(\frac{\lambda_1-\lambda_2}{2})|$ is a generalization of spin reduction factor and $\Delta \phi$ determines the extra phase shift of the oscillation peak position in first harmonic. In the case of both inversion and time-reversal symmetries, $\lambda_1 = -\lambda_2 = \lambda$. The final phase shift $\Delta\phi$ is 0 if $\mathrm{cos}(\lambda)$ is positive ($|\lambda|<0.5\pi$) and $\pi$ if $\mathrm{cos}(\lambda)$ is negative ($|\lambda|>0.5\pi$). If $|\lambda| = 0.5\pi$, the oscillation may not observable because $\mathrm{cos}(\lambda) = 0$.

In 2D systems, the particle number is fixed while chemical potential oscillates around $\mu(B=0)$ with an appreciable amplitude. This oscillation causes a change of the frequency $F$ and the resulting oscillatory behavior is different from 3D cases \cite{Champel2}. But if $\mu>>\hbar \omega_c$ and the reduction factor is small enough, then the oscillation of $\mu$ can be ignored and similar results are obtained \cite{Champel}. It can be calculated by the same method above, where the integration over $k_z$ is omitted (hence, the phase $\pm \pi/4$ disappears). Here, we only give the final expression for the oscillation part of grand canonical potential and magnetization

\begin{equation}
\begin{aligned}
    \frac{\delta\Omega(T,\mu)}{S} &= \sum_{a}\sum_{r} \frac{\hbar \omega_c}{2\pi l_B^2} \frac{1}{\pi^2 r^2} \mathrm{cos}\left[2\pi r(\frac{F}{B}-\gamma_a) \right] R_{T,r} R_{D,r}\\
    &= \sum_{a}\sum_{r} \frac{\hbar \omega_c}{2\pi l_B^2} \frac{1}{\pi^2 r^2} \mathrm{cos}\left[r(l_B^2 A_e+\lambda_a-\phi_M) \right] R_{T,r} R_{D,r} \qquad (2D)
\end{aligned}
\label{omega3D}
\end{equation}

\begin{equation}
\begin{aligned}
    M(T,\mu) &= -\sum_{a}\sum_{r} \frac{\hbar \omega_c}{2\pi l_B^2} \frac{2F}{\pi r B^2} \mathrm{sin}\left[2\pi r(\frac{F}{B}-\gamma_a) \right] R_{T,r} R_{D,r} \\
    &= -\sum_{a}\sum_{r} \frac{\hbar \omega_c}{2\pi l_B^2} \frac{2F}{\pi r B^2} \mathrm{sin}\left[r(l_B^2 A_e+\lambda_a-\phi_M) \right] R_{T,r} R_{D,r} \qquad (2D)
\end{aligned}
\label{M3D}
\end{equation}

\noindent where $S$ is the area of material and reduction factors $R_{T,r}, R_{D,r}$ are the same as before.

Finally, we point out that if one considers higher order terms in the phase $\gamma_a$ (the effective Hamiltonian is expanded to second order in $B$), $\gamma_a$ can depend on the field $B$ \cite{Roth}. If $\gamma_a = \gamma_{a0} + \gamma_{a1} B$ and $\gamma_{a1} << 1/B$, then such a field dependence can be regarded as the change of oscillation frequency with the magnetic field, $F(B)=F(0)-\gamma_{a1} B^2$.

\clearpage

\section{Berry phase in time reversal and inversion symmetric system} \label{PT_berry}
In this section, we discuss the Berry phase in a system with intrinsic spin-orbit coupling, which preserves both time-reversal (TRS) $\mathcal{T}$ and inversion symmetries (IS) $\mathcal{P}$. The kagome lattice model used in the main text has such properties and the analysis here can be directly applied to it. Due to $\mathcal{PT}$ symmetry, every energy band is at least two-fold degenerate so can be labeled by band index $n$ and pseudospin index $s$ ($=\pm 1$). Under time reversal operation, the band index remains the same but pseudospin index s is changed to each other ($s=1\leftrightarrow s=-1$).

First, consider the effect of $\mathcal{T}$. Under time reversal, Bloch states flip pseudospin and momentum but has the same energy (same band index $n$)

\begin{equation}
    \mathcal{T} |u_{n 1}(\boldsymbol{k})\rangle = |u_{n-1}(-\boldsymbol{k})\rangle
\end{equation}

Therefore, the Berry connection for different pseudospin components has the relation

\begin{equation}
    \boldsymbol{A}_{n-1}(-\boldsymbol{k}) = i\langle u_{n-1}(-\boldsymbol{k}) |\nabla_{-k}| u_{n-1}(-\boldsymbol{k}) \rangle = -i\langle \mathcal{T} u_{n 1}(\boldsymbol{k}) |\nabla_{k}| \mathcal{T} u_{n 1}(\boldsymbol{k}) \rangle = -i\langle \nabla_{k} u_{n 1}(\boldsymbol{k}) | u_{n 1}(\boldsymbol{k}) \rangle = i\langle u_{n 1}(\boldsymbol{k}) |\nabla_{k}| u_{n 1}(\boldsymbol{k}) \rangle = \boldsymbol{A}_{n 1}(\boldsymbol{k})
\label{A}
\end{equation}

The Berry phase along the loop $\mathcal{C}$ for one pseudospin is therefore opposite to the Berry phase of the inversion loop $\overline{\mathcal{C}}$ for opposite pseudospin

\begin{equation}
    \phi_{n 1}(\mathcal{C}) = \int_{\mathcal{C}} \boldsymbol{A}_{n 1}(\boldsymbol{k}) \cdot d \boldsymbol{k} = \int_{\mathcal{C}} \boldsymbol{A}_{n-1}(-\boldsymbol{k}) \cdot d \boldsymbol{k} = -\int_{\mathcal{C}} \boldsymbol{A}_{n-1}(-\boldsymbol{k}) \cdot d (-\boldsymbol{k}) = -\int_{\overline{\mathcal{C}}} \boldsymbol{A}_{n-1}(\boldsymbol{k}) \cdot d \boldsymbol{k} = - \phi_{n-1}(\overline{\mathcal{C}})
\label{gamma}
\end{equation}

Next, we consider the role of inversion symmetry. Under inversion operation, Bloch states reverse momentum but preserves pseudospin and energy

\begin{equation}
    \mathcal{P} |u_{n s}(\boldsymbol{k})\rangle = |u_{n s}(-\boldsymbol{k})\rangle
\end{equation}

The Berry connection for the same pseudospin at opposite positions has the relation

\begin{equation}
    \boldsymbol{A}_{n s}(-\boldsymbol{k}) = i\langle u_{n s}(-\boldsymbol{k}) |\nabla_{-k}| u_{n s}(-\boldsymbol{k}) \rangle = -i\langle \mathcal{P} u_{n s}(\boldsymbol{k}) |\nabla_{k}| \mathcal{P} u_{n s}(\boldsymbol{k}) \rangle = -i\langle u_{n s}(\boldsymbol{k}) |\nabla_{k}| u_{n s}(\boldsymbol{k}) \rangle = -\boldsymbol{A}_{n s}(\boldsymbol{k})
\end{equation}

So the Berry phase along the loop $\mathcal{C}$ for one pseudospin now is the same for the inversion loop $\overline{\mathcal{C}}$ with the same pseudospin

\begin{equation}
    \phi_{n s}(\mathcal{C}) = \int_{\mathcal{C}} \boldsymbol{A}_{n s}(\boldsymbol{k}) \cdot d \boldsymbol{k} = -\int_{\mathcal{C}} \boldsymbol{A}_{n s}(-\boldsymbol{k}) \cdot d \boldsymbol{k} = \int_{\mathcal{C}} \boldsymbol{A}_{n s}(-\boldsymbol{k}) \cdot d (-\boldsymbol{k}) = \int_{\overline{\mathcal{C}}} \boldsymbol{A}_{n s}(\boldsymbol{k}) \cdot d \boldsymbol{k} =  \phi_{n s}(\overline{\mathcal{C}})
\end{equation}

If the system has both $\mathcal{T}$ and $\mathcal{P}$, combining the results above, we obtain that Berry phases around the same loop for different pseudospin components are opposite

\begin{equation}
    \begin{aligned}
    \boldsymbol{A}_{n s}(\boldsymbol{k}) &= \boldsymbol{A}_{n-s}(-\boldsymbol{k}) = -\boldsymbol{A}_{n-s}(\boldsymbol{k}) \\
    \phi_{n s}(\mathcal{C}) &= - \phi_{n-s}(\overline{\mathcal{C}}) =  -\phi_{n-s}(\mathcal{C})
    \end{aligned}
\label{Berryphase}
\end{equation}

\begin{figure}
\centering
\includegraphics[width=0.5\linewidth]{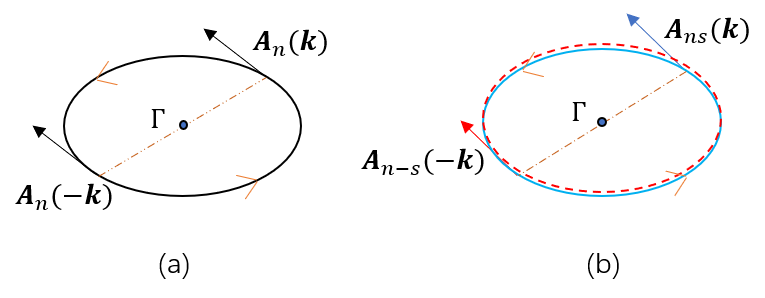}
\caption{\label{FigS6}
The Berry connection for the loop around $\Gamma$ point, (a) for a spinless case whose Berry phase is zero, and (b) for a spinful case with two spin components. The blue line represents spin $s$ and the red line represents spin $-s$. Their Berry phases are opposite to each other.
}
\end{figure}

The Berry phase with spin taken into consideration is very different from that of the spinless case. For example, consider a Fermi surface around the Brillouin zone center ($\Gamma$ point, Fig. \ref{FigS6}) and assume only time-reversal symmetry. In the spinless case, band energy satisfies $\varepsilon(\boldsymbol{k})=\varepsilon(-\boldsymbol{k})$, the trajectory of such a Fermi surface must be symmetric with respect to $\Gamma$, i.e., $\mathcal{C}=\overline{\mathcal{C}}$. Equations \eqref{A} and \eqref{gamma} become

\begin{equation}
    \begin{aligned}
        \boldsymbol{A}_{n}(\boldsymbol{k}) &= \boldsymbol{A}_{n}(-\boldsymbol{k}) \\
        \phi_{n}(\mathcal{C}) &= - \phi_{n}(\overline{\mathcal{C}}) = 0
    \end{aligned}
\end{equation}

\noindent The Berry phase around $\Gamma$ point must be zero. This argument is also applied to Fermi surfaces surrounding any other time-reversal invariant momenta. But if the spin is added, then one can only draw the conclusion from \eqref{gamma} that Berry phases of two pseudospins with $\mathcal{T}$-correlated loops $\mathcal{C}$ and $\overline{\mathcal{C}}$ are opposite.

Finally, for a system with both $\mathcal{T}$ and $\mathcal{P}$, not only the Berry phases for the two pseudospin channels with the same loop $\mathcal{C}$ \eqref{Berryphase} are opposite, but also the total phases (i.e., the generalized Berry phase $\lambda_a$, composed of Berry phase, orbital phase, and Zeeman phase) of two pseudospin channels are opposite \cite{Aris2018PRX}. As a result, the extra phase shift $\Delta \phi$ \eqref{spinsum} in the quantum oscillation can only be $0$ or $\pi$, depending on whether $\lambda_a$ is smaller or larger than $0.5\pi$. Hence, in general, the $\pi$ phase determined from the oscillation peak position can't serve as evidence of a Dirac point but may solely come from the system's symmetry\cite{Aris2018PRX}.

\clearpage

\section{Berry phase and orbital phase in Dirac and BHZ model} \label{Dirac_oscillation}

First, we apply the above results to the 2D massive Dirac model. It can describe graphene with SOC taking into account or the cross section in k space of Weyl semimetal if the mass is zero. Since the band gap opens at the Dirac point, the Berry phase is usually no longer $\pi$. But the orbital moment phase can compensate the Berry phase and their sum can be $\pi$ independent of the mass $\Delta$ of Dirac fermion\cite{Fuchs_2010}.

To be more precise, consider the following two band Dirac Hamiltonian with Semenoff mass $\Delta$

\begin{equation}
    H_s(k) = \hbar v_F(k_x \sigma_x + k_y \sigma_y) + \Delta \sigma_z
\label{Dirac}
\end{equation}

\noindent In the graphene case, $\Delta=s\lambda_{SO}$ with $\lambda_{SO}>0$ represents SOC strength and $s=\pm 1$ represents different spin indices. The Hamiltonian gives rise two bands with energy $\pm\varepsilon(\boldsymbol{k})=\pm\sqrt{\hbar^2 v_F^2 \boldsymbol{k}^2 + \Delta^2}$. Since band structure is non-degenerate, three contributions to $\lambda_a$ can be calculated separately. The Berry phase of the conduction band for a loop with Fermi wave vector $k_F$ around the Dirac point is

\begin{equation}
    \phi_B(k_F) = \int_{|\boldsymbol{k}|=k_F} \boldsymbol{A}(\boldsymbol{k}) \cdot d \boldsymbol{k} = \int_{|\boldsymbol{k}|<k_F} d k_x d k_y F_{x y}(\boldsymbol{k}) = -\pi(1-\frac{\Delta}{\varepsilon(k_F)})
\end{equation}

\noindent where $F_{x y}=-\Delta/(\hbar^2 v_F^2 k_F^2 + \Delta^2)^{3/2}$ is the Berry curvature $\Omega$. The Berry phase is near zero at the band bottom and approaches $-\pi$ when $E_F=\varepsilon(k_F)$ is much larger than the gap $\Delta$. In this two-band model with particle-hole symmetry, the orbital magnetic moment is proportional to Berry curvature $M(\boldsymbol{k})=\left[e/(\hbar c)\right]\varepsilon(\boldsymbol{k})\Omega$. Hence, the orbital moment phase \eqref{phi_R} is

\begin{equation}
    \phi_R(k_F) = l_B^2\oint MB \frac{|d\boldsymbol{k}|}{v^{\perp}}=\oint \varepsilon\Omega \frac{|d\boldsymbol{k}|}{v^{\perp}}=\varepsilon\frac{d\phi_B}{d\varepsilon}
\end{equation}

\noindent where $v^{\perp}$ is the group velocity component perpendicular to $B$. If the Zeeman phase $\phi_Z$ can be ignored or extracted as a spin reduction factor, $\lambda_a$ is $-\pi$ independent of $\Delta$.

\begin{equation}
    \phi_B + \phi_R=\phi_B+\varepsilon\frac{d\phi_B}{d\varepsilon}=\frac{d(\varepsilon\phi_B)}{d\varepsilon}=-\pi
\label{pi_Berry}
\end{equation}

\begin{figure}[htbp]
\centering
\subfigure[]
{
\includegraphics[width=0.45\linewidth]{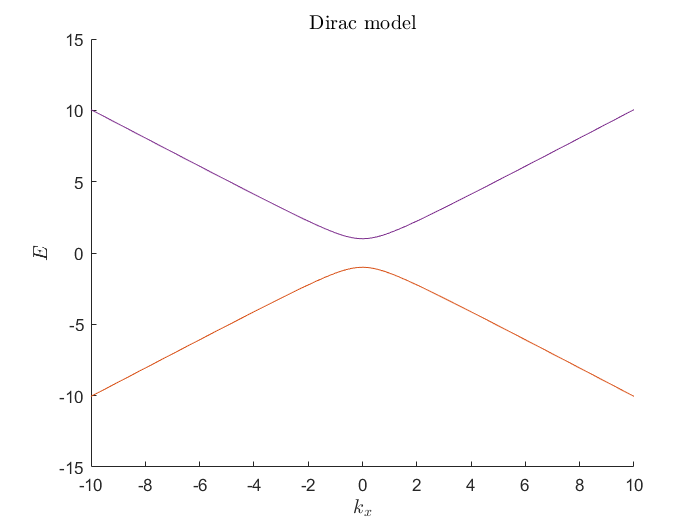}
}
\subfigure[]
{
\includegraphics[width=0.45\linewidth]{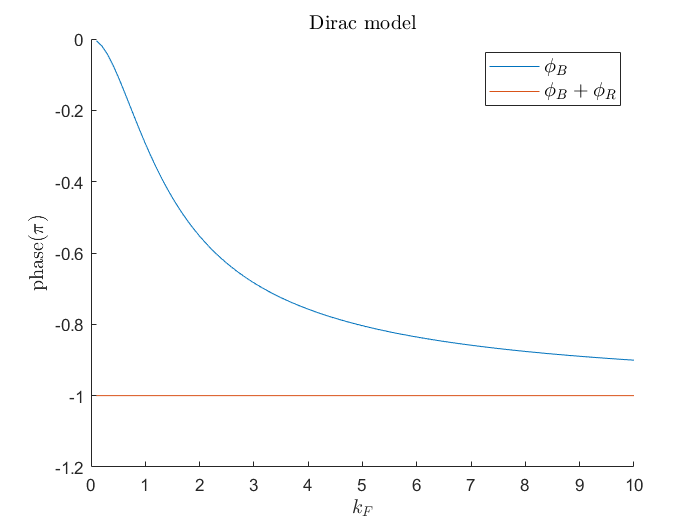}
}
\subfigure[]
{
\includegraphics[width=0.45\linewidth]{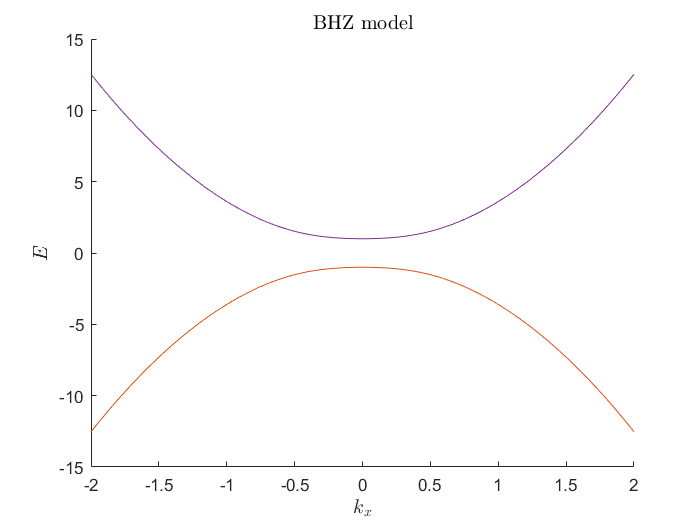}
}
\subfigure[]
{
\includegraphics[width=0.45\linewidth]{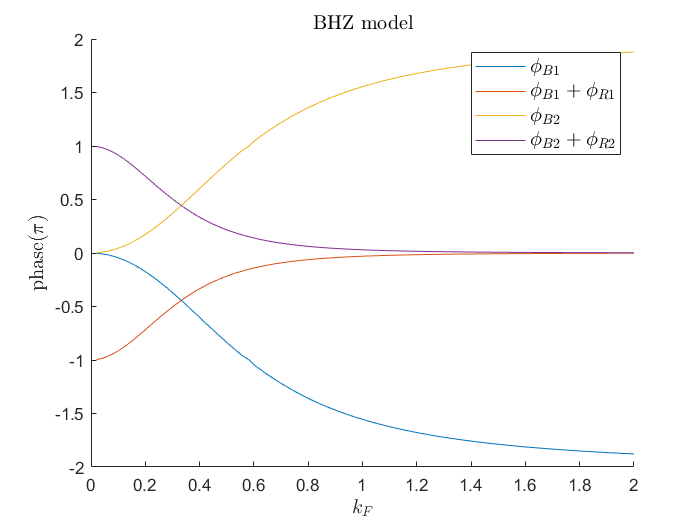}
}
\caption{\label{FigS7}
(a) Band energy and (b) Berry phase and the phase sum of gapped Dirac model \eqref{Dirac}. The parameters are $\hbar v=1$ and Semenoff mass $\Delta=1$. (b) Band energy in $k_z=0$ mirror plane and (c) Berry phase and the phase sum of BHZ model \eqref{BHZ}. The parameters are $A=B=3$ and $M_0=\tilde{A}=1$ which is in $\mathbb{Z}_2$ topological insulator region.
}
\end{figure}

The numerically calculated Berry phase $\phi_B$ and the phase sum $\phi_B+\phi_R$ are shown in Fig. \ref{FigS7}, which is consistent with \eqref{pi_Berry}. Hence, even if the Dirac crossing is gapped by SOC, the oscillation phase can still manifest itself as $\pi$. It can be attributed to the pure Berry phase only when the mass is zero. In the case of Graphene, the SOC gap $\Delta$ is much smaller than typical Fermi energy, which leads to an almost zero $\phi_R$ and $\pi$ Berry phase (the far right of Fig. \ref{FigS7} (b)).

Next, we consider the 3D BHZ model which represents the SOC-induced band anti-crossing

\begin{equation}
H(\mathbf{k})=\epsilon(\mathbf{k}) \cdot I+A(k_x \sigma_x s_z - k_y \sigma_y) + M(\mathbf{k})\sigma_z + \tilde{A}k_z \sigma_x s_x=\epsilon(\mathbf{k}) \cdot I+\left(\begin{array}{cccc}
M(\mathbf{k}) & A k_{+} & 0 & \tilde{A} k_z \\
A k_{-} & -M(\mathbf{k}) & \tilde{A} k_z & 0 \\
0 & \tilde{A} k_z & M(\mathbf{k}) & -A k_{-} \\
\tilde{A} k_z & 0 & -A k_{+} & -M(\mathbf{k})
\end{array}\right)
\label{BHZ}
\end{equation}

\noindent where $\epsilon(\mathbf{k})$ contributes only to the band energy and can be set as zero. $k_{\pm}=k_x\pm ik_y$ and $M(\mathbf{k}) = M_0-B(k_x^2+k_y^2+k_z^2)$ is a $k_z$ dependent mass. This model is a trivial insulator when $M_0/B<0$ while is a $\mathbb{Z}_2$ topological insulator when $M_0/B>0$. It has several important symmetries: time reversal and inversion symmetry which ensure energy bands are two-fold degenerate, particle-hole symmetry, and mirror symmetry with respect to $k_z=0$ plane. The last two symmetries can be mathematically represented as

\begin{equation}
    (\sigma_x s_z) H(\mathbf{k}) (\sigma_x s_z)^{-1} = -H(\mathbf{k}) \qquad s_z H(\mathbf{k}^{\perp},k_z) s_z^{-1} = H(\mathbf{k}^{\perp},-k_z)
\end{equation}

Due to the mirror symmetry, the extremal cyclotron orbit lies on the $k_z=0$ plane and we focus on this orbit in the conduction band. This orbit is a circle around $\Gamma$ point with Fermi wave vector $k_F$ and has two-fold degeneracy. The band energy in this mirror symmetry plane and the magnetic oscillation phase associated with this extremal orbit are shown in Fig. \ref{FigS7}, which has an opposite magnetic oscillation phase for degenerate bands due to $\mathcal{PT}$ symmetry. The Berry phase grows continuously from 0 to $2\pi$ while the phase sum $\phi_B+\phi_R$ decreases from $\pi$ to 0 as $k_F$ is larger. When the energy is close to the band edge, the total oscillation phase is close to $\pi$ if spin contribution $\phi_Z$ can be ignored. However, this phase originates solely from the orbital moment and the Berry phase is nearly zero instead.

\clearpage

\begin{table}
\centering
\caption{\label{Table_phi_40} Berry phase $\phi_B$, orbital phase $\phi_R$ and Zeeman phase $\phi_Z$ of quantum orbits of 2$\times$2$\times$2 CDW at $-$40 meV calculated by using the expression \eqref{Aris propagator}. Under SOC, the orbital and spin magnetic moment-related phases can only be calculated together with the Berry phase, and ($\phi_B$+$\phi_R$+$\phi_Z$) is the generalized Berry phase ($\lambda$) in the main text. Frequency (Freq.) is unit of T. All phases are in unit of $\pi$. For the 912 T, we could not get converged results for $\phi_B$+$\phi_R$ and $\phi_B$+$\phi_R$+$\phi_Z$. Notice that the numerical convergence of $\lambda_a$ for the quantum orbit of 912 T at --40 meV is not as good as others.
}
\renewcommand\arraystretch{1.2}
\begin{ruledtabular}
\begin{tabular}{ccccccccccc}
  \multirow{2}{*}{Frequency} & \multicolumn{3}{c}{no SOC} & & \multicolumn{3}{c}{SOC}\\
  \cline{2-4}
  \cline{6-8}
     & $\phi_B$ & $\phi_R$ & $\phi_Z$  &  & $\phi_B$ & $\phi_B$+$\phi_R$ & $\phi_B$+$\phi_R$+$\phi_Z$ \\
    \hline
    \uline{13}      & 1 & 0   & 0.27 && 0.83 & 0.92   & 0.68 \\
    20              & 0 & 0   & 0.44 && 0.00 & 0.04   & 0.38 \\
    \uline{30}      & 1 & 0   & 0.30 && 0.85 & 0.83   & 0.83 \\
    \uline{77}      & 0 & 0   & 0.68 && 0.51 & 0.35   & 0.92 \\
    \uline{98}      & 0 & 0   & 0.45 && 0.36 & 0.49   & 0.87 \\
    \uline{152}     & 0 & 0   & 0.57 && 0.14 & 0.60   & 0.98 \\
    202             & 0 & 0   & 0.83 && 0.56 & 0.53   & 0.37 \\
    \uline{230}     & 0 & 0   & 0.50 && 0.51 & 0.54   & 0.84 \\
    609             & 0 & 0   & 0.30 && 0.17 & 0.18   & 0.48 \\
    \uline{730}     & 0 & 0   & 0.38 && 0.63 & 0.66   & 0.92 \\
    734             & 0 & 0   & 0.22 && 0.90 & 0.33   & 0.22 \\
    894             & 0 & 0   & 0.47 && 0.61 & 0.66   & 0.19 \\
    \uline{898}     & 0 & 0   & 0.42 && 0.23 & 0.34   & 0.71 \\
    912             & 0 & 0   & 0.22 && 0.45 & 0.97   & 0.31 \\
    1067            & 0 & 0   & 0.89 && 0.06 & 0.82   & 0.33 \\
    1445            & 0 & 0   & 0.50 && 0.79 & 0.60   & 0.26 \\
\end{tabular}
\end{ruledtabular}
\end{table}

\begin{table}
\centering
\caption{\label{Table_phi_85} Similar to Table \ref{Table_phi_40} but at --85 meV.
}
\renewcommand\arraystretch{1.2}
\begin{ruledtabular}
\begin{tabular}{ccccccccccccc}
  \multirow{2}{*}{Frequency} & \multicolumn{3}{c}{no SOC} & & \multicolumn{3}{c}{SOC}\\
  \cline{2-4}
  \cline{6-8}
     & $\phi_B$ & $\phi_R$ & $\phi_Z$  &  & $\phi_B$ & $\phi_B$+$\phi_R$ & $\phi_B$+$\phi_R$+$\phi_Z$ \\
    \hline
    \uline{5}   &   &    &      && 0.14 & 0.79   & 0.93 \\
    \uline{14}  & 1 & 0  & 0.26 && 0.99 & 0.98   & 0.77 \\
    \uline{33}  & 0 & 0  & 0.23 && 0.85 & 0.59   & 0.84 \\
    40          & 0 & 0  & 0.15 && 0.82 & 0.64   & 0.44 \\
    \uline{51}  & 0 & 0  & 0.21 && 0.65 & 0.34   & 0.52 \\
    76          & 0 & 0  & 0.33 && 0.72 & 0.25   & 0.02 \\
    \uline{99}  & 0 & 0  & 0.49 && 0.72 & 0.62   & 0.89 \\
    \uline{224} & 0 & 0  & 0.64 && 0.01 & 0.07   & 0.57 \\
    429         & 0 & 0  & 0.59 && 0.57 & 0.34   & 0.06 \\
    605         & 1 & 0  & 0.70 && 0.98 & 0.92   & 0.38 \\
    617         & 1 & 0  & 0.71 && 0.99 & 0.93   & 0.35 \\
    672         & 0 & 0  & 0.25 && 0.76 & 0.02   & 0.11 \\
    \uline{735} & 0 & 0  & 0.35 && 0.18 & 0.20   & 0.55 \\
    \uline{765} & 0 & 0  & 0.93 && 0.71 & 0.08   & 0.60 \\
    940         & 0 & 0  & 0.48 && 0.30 & 0.53   & 0.07 \\
    1267        & 0 & 0  & 0.76 && 0.84 & 0.98   & 0.25 \\
    1293        & 0 & 0  & 0.03 && 0.67 & 0.57   & 0.08 \\
\end{tabular}
\end{ruledtabular}
\end{table}

\clearpage
%